\documentclass[fleqn,usenatbib]{mnras}

\usepackage[T1]{fontenc}

\DeclareRobustCommand{\VAN}[3]{#2}
\let\VANthebibliography\thebibliography
\def\thebibliography{\DeclareRobustCommand{\VAN}[3]{##3}\VANthebibliography}

\usepackage{graphicx}	% Including figure files
\usepackage{amsmath}	% Advanced maths commands
\usepackage{amssymb}	% Extra maths symbols
\usepackage{newtxtext,newtxmath}
\usepackage[separate-uncertainty=true,tight-spacing=true]{siunitx}
\usepackage{multirow}
\usepackage[labelfont=bf,labelsep=period]{caption}
\usepackage{bm}

\newcommand{\fig}[1]{Fig.~\ref{#1}}

\DeclareSIUnit{\solarM}{M_{\odot}}
\DeclareSIUnit{\solarL}{L_{\odot}}
\DeclareSIUnit{\parsec}{pc}
\DeclareSIUnit{\year}{yr}
\DeclareSIUnit{\erg}{erg}
\DeclareSIUnit\angstrom{\text {Å}}

\title[Evolution of the Dual AGN in Mrk 266]{
Evolution of the Dual AGN in Mrk 266: A Young AGN and a Rotation Dominated Disk in the SW Nucleus
}

\author[Ruby et al.]{
    Mason Ruby,$^{1}$\thanks{E-mail: \href{mailto:mdruby@memphis.edu}{mdruby@memphis.edu}} Francisco Müller-Sánchez,$^{1}$ Julia M.\ Comerford,$^{2}$ Daniel Stern,$^{3}$ Sabrina L.\ Cales,$^{4,5}$ \newauthor Fiona Harrison,$^{6}$ Matthew A.\ Malkan,$^{7}$ George C.\ Privon,$^{8,9}$ Ezequiel Treister$^{10}$
\\
$^{1}$Department of Physics and Materials Science, The University of Memphis, Memphis, TN 38152, USA \\
$^{2}$Department of Astrophysical and Planetary Sciences, University of Colorado, Boulder, CO 80309, USA \\
$^{3}$Jet Propulsion Laboratory, California Institute of Technology, 4800 Oak Grove Drive, Mail Stop 264-789, Pasadena, CA 91109, USA \\
$^{4}$Yale Center for Astronomy and Astrophysics, Physics Department, Yale University, New Haven, CT 06511, USA \\
$^{5}$Department of Astronomy, University of Concepcion, Concepcion, Chile \\
$^{6}$Cahill Center for Astronomy and Astrophysics, California Institute of Technology, Pasadena, CA 91125, USA \\
$^{7}$Department of Physics and Astronomy University of California, Los Angeles Los Angeles, CA 90095-1547, USA \\
$^{8}$National Radio Astronomy Observatory, 520 Edgemont Road, Charlottesville, VA 22903, USA \\ 
$^{9}$Department of Astronomy, University of Florida, P.O. Box 112055, Gainesville, FL 32611, USA \\
$^{10}$Instituto de Astrofísica, Facultad de Física, Pontificia Universidad Católica de Chile, Santiago, Chile \\
}

\date{Accepted XXX. Received YYY; in original form ZZZ}

\pubyear{2023}

\begin{document}
\label{firstpage}
\pagerange{\pageref{firstpage}--\pageref{lastpage}}
\maketitle

\begin{abstract}
    Dual active galactic nuclei (AGN) offer a unique opportunity to probe the relationship between super massive black holes (SMBH) and their host galaxies as well as the role of major mergers in triggering AGN activity. The confirmed dual AGN Mrk 266 has been studied extensively with multi-wavelength imaging. Now, high spatial resolution IFU spectroscopy of Mrk 266 provides an opportunity to probe the kinematics of both the merger event and AGN feedback. We present for the first time high spatial resolution kinematic maps for both nuclei of Mrk 266 obtained with the Keck OSIRIS IFU spectrograph, utilizing adaptive optics to achieve a resolution of $0.31\arcsec$ and $0.20\arcsec$ for the NE and SW nuclei, respectively. Using the $M_\text{BH}$\nobreakdash--$\sigma_*$ relation for mergers, we infer a SMBH mass of approximately \qty{7e7}{\solarM} for the southwestern nucleus. Additionally, we report that the molecular gas kinematics of the southwestern nucleus are dominated by rotation rather than large-scale chaotic motions.  The southwest nucleus also contains both a circumnuclear ring of star formation from which an inflow of molecular gas is likely fueling the AGN and a compact, AGN-dominated outflow of highly ionized gas with a timescale of approximately \qty{2}{\mega\year}, significantly shorter than the timescale of the merger.  The northeastern nucleus, on the other hand, exhibits complex kinematics related to the merger, including molecular gas that appears to have decoupled from the rotation of the stars.  Our results suggest that while the AGN activity in Mrk 266 was likely triggered during the merger, AGN feeding is currently the result of processes internal to each host galaxy, thus resulting in a strong asymmetry between the two nuclei.
\end{abstract}


%%%%%%%%%%%%%%%%%%%%%%%%%%%%%%%%%%%%%%%%%%%%%%%%%%

%%%%%%%%%%%%%%%%% BODY OF PAPER %%%%%%%%%%%%%%%%%%

\section{Introduction}

It is now widely understood that all massive galaxies contain a central super massive black hole (SMBH) with mass ranging from  $10^{6}-10^{10}\,M_{\odot}$ \citep{Ferrarese_and_Ford}.  While the gravitational sphere of influence for SMHBs is on the order of $1-100$ pc, there are many correlations between the SMBH mass and properties of the host galaxy at kpc scales, most notably stellar velocity dispersion of the bulge ($M_{\text{BH}}-\sigma_*$) \citep{Ferrarese_and_Merrit,Gebhardt,McConell_and_Ma}.  These correlations suggest a connection between the evolution of SMBHs and their host galaxies; however, the details of this connection are yet to be fully understood \citep{Kormandy_and_Ho}.  Also, massive galaxies are known to be the product of major mergers \citep[and references therein]{Conselice2008,ConseliseRev,Bluck} which can trigger active galactic nuclei (AGN) \citep{Ellison,Goulding,Gao} in which the SMBH accretes matter, generating bolometric luminosities on the order of $10^{10}-10^{14}\,L_{\odot}$ \citep{Woo_and_Urry}.  Dual AGN, mergers with kpc-scale separations in which both nuclei are accreting, are the quintessential laboratory for probing the SMBH-galaxy connection, exhibiting several complex processes that are rarely detected simultaneously, e.g. vigorous star formation, multiple AGN, outflowing winds of ionized gas, and tidal distortions.
        \begin{figure*}
    % 	    % To include a figure from a file named example.*
    % 	    % Allowable file formats are eps or ps if compiling using latex
    % 	    % or pdf, eps, jpg if compiling using pdf latex
 	    \includegraphics[width=0.9\textwidth]{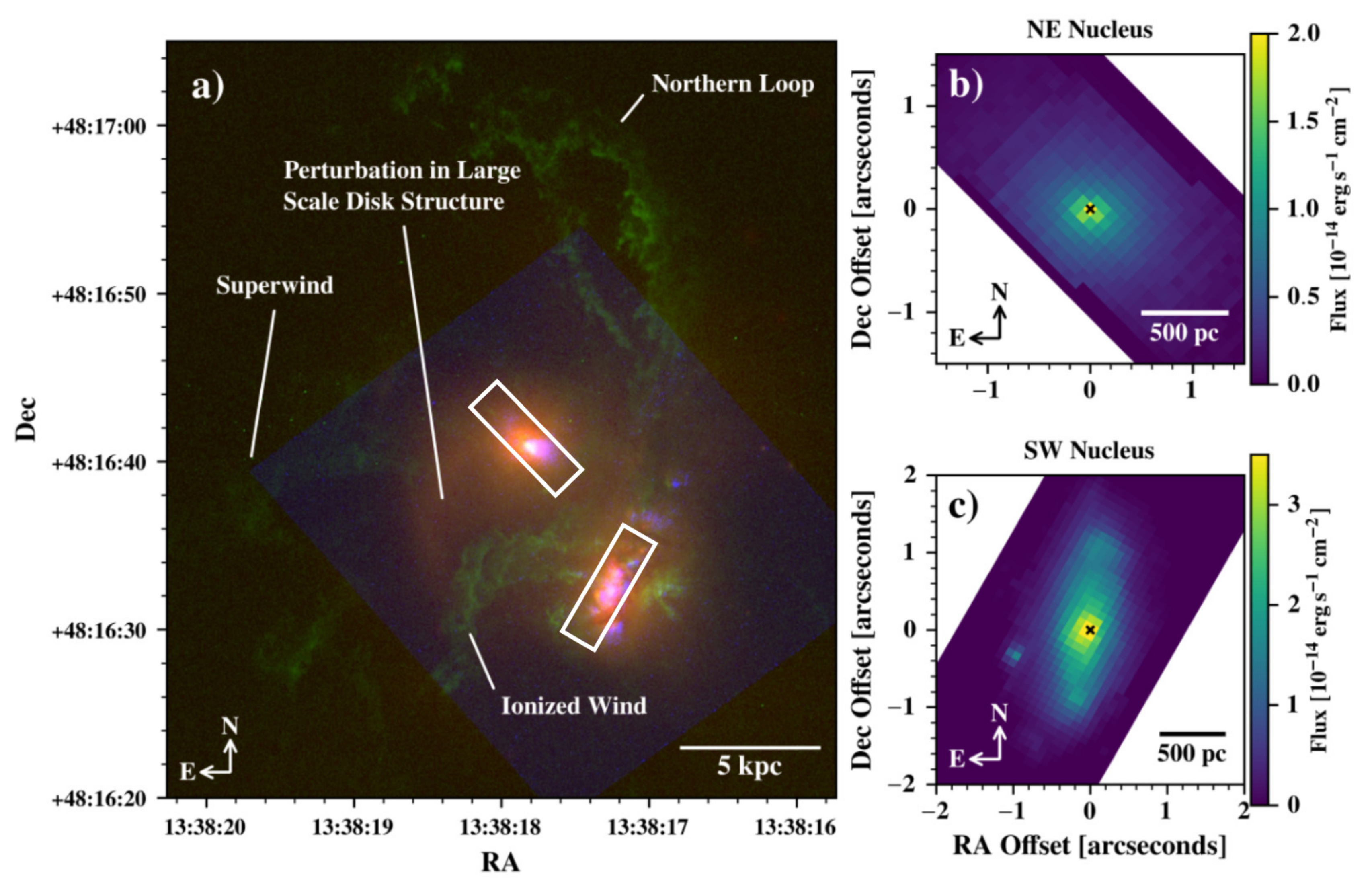}
         \caption{False color composite image of Mrk 266 using data from \textit{HST}. The blue channel shows the F330W filter (UV continuum), green F673N ($\text{H}\alpha$ emission), and red F814W (red continuum).  The white rectangles indicate the approximate FOV of OSIRIS.  The UV continuum shows regions of star formation in the central region of the SW nucleus.  The $\text{H}\alpha$ emission shows large tidal features to the north and east as well as a superwind to the east \citep{Mazzarella2012}.  The red continuum shows tidal perturbations in the large-scale structure of the NE nucleus and strong dust obscuration in the SW.  The separation of the two nuclei is approximately $10\arcsec$ or \qty{6}{\kilo\parsec}. (\textit{b}) \textit{K}-band integrated flux map of the NE nucleus from Keck/OSIRIS adaptive-optics-assisted IFU observations. (\textit{c}) The same as (\textit{b}) for the SW nucleus.}
         \label{fig:MRK266-fcc}
     \end{figure*}
    \begin{table*}
        \centering
        \caption{Emission line fluxes for all detected emission lines in the SW and NE nuclei.  Column $(3)$ reports the flux for the entire FOV, $(4)$ the flux in a $1\arcsec$ diameter aperture centered on the AGN, and $(5)$ the flux in a $0.5\arcsec$ diameter aperture centered on the AGN.  Dashes indicate the amplitude of the emission line was below $2\sigma$, and stellar absorption lines are reported with equivalent width in \unit{\angstrom} rather than flux. $1\sigma$ errors are reported.}
        \label{tab:EmissionLineFlux}
        \renewcommand{\arraystretch}{1.3}
        \begin{tabular}{c c c c c}
            \hline
            \multirow{2}{*}{Emission Line} &  Rest Wavelength & \multicolumn{1}{c}{Flux (Full FOV)} & \multicolumn{1}{c}{Flux ($1\arcsec$)} & \multicolumn{1}{c}{Flux ($0.5\arcsec$)} \\
             & $[\unit{\angstrom}]$ & \multicolumn{1}{c}{$[10^{-15}\,\text{erg}\,\text{s}^{-1}\,\text{cm}^{-2}]$} & \multicolumn{1}{c}{$[10^{-15}\,\text{erg}\,\text{s}^{-1}\,\text{cm}^{-2}]$} & \multicolumn{1}{c}{$[10^{-15}\,\text{erg}\,\text{s}^{-1}\,\text{cm}^{-2}]$} \\
            \hline
            \multicolumn{5}{c}{SW Nucleus}\\
            \hline 
            $\text{Br}\delta$* &   $19451$   & \textemdash  & $0.87\pm0.32$ & $0.31\pm0.10$  \\
            $[\ion{Si}{vi}]$* & $19641$ &     \textemdash      & $1.18\pm0.26$  & $1.24\pm0.11$  \\
            $\text{H}_2\,[2\text{--}1\,S(4)]$ & $20041$ & $10.2\pm1.4$  & $0.64\pm0.14$  & $0.182\pm0.048$   \\
            $\text{H}_2\,[1\text{--}0\,S(2)]$ & $20338$ & $4.97\pm0.60$  & $1.170\pm0.069$  & $0.393\pm0.025$ \\
            $\ion{He}{i}$  & $20587$ & $5.4\pm1.1$  & $1.094\pm0.089$  & $0.374\pm0.029$ \\
            $\text{H}_2\,[2\text{--}1\,S(3)]$ & $20735$ & $2.50\pm0.42$  & $0.258\pm0.057$   &  $0.096\pm0.032$   \\
            $\text{H}_2\,[1\text{--}0\,S(1)]$ & $21218$ & $10.0\pm1.4$  & $2.36\pm0.11$  & $0.840\pm0.039$  \\
            $\text{Br}\gamma$     & $21661$ & $24.1\pm2.5$  & $3.83\pm0.10$ & $1.239\pm0.034$  \\
            $\text{H}_2\,[1\text{--}0\,S(0)]$ & $22235$ & $5.46\pm0.95$  & $0.91\pm0.15$  & $0.371\pm0.049$  \\
            $^{12}\text{CO}\,[2\text{--}0]$** & $22935$ & $12.8\pm3.3$ \unit{\angstrom}  & $9.7\pm1.2$ \unit{\angstrom} & $9.5\pm1.2$ \unit{\angstrom}  \\
            Integrated \textit{K}-band & \textemdash & \multicolumn{1}{c}{12.24 mag} & \multicolumn{1}{c}{12.55 mag} & \multicolumn{1}{c}{13.56 mag} \\
            \hline 
            \multicolumn{5}{c}{NE Nucleus}\\
            \hline
            $\text{Br}\delta$ &   $19451$   & $1.55\pm0.34$  & $1.122\pm0.066$ & $0.467\pm0.026$  \\
            $\text{H}_2\,[1\text{--}0\,S(3)]$ & $19576$ & $7.25\pm0.40$  & $3.962\pm0.091$ & $1.338\pm0.032$ \\
            $\text{H}_2\,[2\text{--}1\,S(4)]$ & $20041$ & $0.95\pm0.10$   & $0.297\pm0.050$ & $0.074\pm0.019$\\
            $\text{H}_2\,[1\text{--}0\,S(2)]$ & $20338$ & $3.33\pm0.10$  & $1.662\pm0.033$ & $0.492\pm0.012$  \\
            $\ion{He}{i}$  & $20587$ & $1.84\pm0.13$  & $1.242\pm0.043$ & $0.448\pm0.018$  \\
            $\text{H}_2\,[2\text{--}1\,S(3)]$ & $20735$ &  $0.52\pm0.15$     & $0.394\pm0.041$ & $0.136\pm0.017$  \\
            $\text{H}_2\,[1\text{--}0\,S(1)]$ & $21218$ & $7.29\pm0.18$  & $4.013\pm0.071$ & $1.466\pm0.029$ \\
            $\text{Br}\gamma$     & $21661$ & $4.57\pm0.16$  & $2.706\pm0.072$ & $1.043\pm0.030$ \\
            $\text{H}_2\,[3\text{--}2\,S(3)]$ & $22014$ &   $0.85\pm0.15$  & $0.200\pm0.030$  &  $0.055\pm0.021$ \\
            $\text{H}_2\,[1\text{--}0\,S(0)]$ & $22235$ & $2.08\pm0.18$ & $1.159\pm0.068$ & $0.417\pm0.025$  \\
            $\text{H}_2\,[2\text{--}1\,S(1)]$ & $22477$ & $1.86\pm0.18$  & $0.570\pm0.055$  & $0.153\pm0.022$  \\
            $^{12}\text{CO}\,[2\text{--}0]$** & $22935$ & $5.3\pm1.3$ \unit{\angstrom} & $5.85\pm0.86$ \unit{\angstrom} & $6.9\pm1.1$ \unit{\angstrom} \\
            Integrated \textit{K}-band & \textemdash & \multicolumn{1}{c}{12.68 mag} & \multicolumn{1}{c}{13.62 mag} & \multicolumn{1}{c}{14.87 mag} \\
            \hline
            \multicolumn{5}{p{5in}}{*The $\text{Br}\delta$ and $[\ion{Si}{vi}]$ emission lines in the SW nucleus were affected by telluric absorption, particularly when integrating the full FOV.} \\
            \multicolumn{5}{p{5in}}{**The strength of the $^{12}\text{CO}\,[2\text{--}0]$ stellar absorption bandhead is reported as its equivalent width in \unit{\angstrom}.} 
        \end{tabular}
    \end{table*}

Simulations have shown that galaxy mergers can cause gas in the interstellar medium to lose angular momentum and flow toward the nuclear regions where it is available for AGN accretion and can trigger a burst of galaxy-wide star formation \citep{Mihos_and_Hernquist,Hopkins_and_Hernquist,Springel,Quai}.  However, observational studies have shown mixed results regarding the importance of mergers in AGN feeding compared to secular processes such as axisymmetric perturbations in the disk \citep{Davies,Smethurst2019,Gao, Stemo}.  Additionally, though galaxy mergers are prevalent and thought to trigger AGN, detections of dual AGN are rare, with $\sim\!30$ confirmed so far \citep{Satyapal,Treister2012,Koss2012}.  The apparent dearth of dual AGN, despite several systematic searches using SDSS spectra \citep{Liu2010,Comerford} as well as radio and X-ray data \citep{Tingay_and_Warth,Liu2013,Fu,Gabanyi}, may be due to heavy obscuration of the nuclear regions and short timescales of merging galaxies \citep{DeRosa}; however, there may also be physical processes occurring in dual AGN that increase the probability of simultaneous SMBH growth \citep{Solanes,Ellison2011,Capelo,Koss2012}.  The resulting AGN feedback, in the form of kpc scale winds of ionized gas \citep{Mullaney,Rupke,Muller_Sanchez2011,Muller_Sanchez2016}, may then quench both star formation rates (SFR) and further AGN activity; however, counter examples of AGN-triggered star formation complicate this picture \citep{Juneau,Schutte}.  

    \subsection{Mrk 266: A Confirmed Dual AGN in a Mid-stage Merger}

    %We present results from Keck$/$OSIRIS adaptive optics (AO) assisted integral field unit (IFU) spectroscopy of the confirmed dual AGN Mrk 266 \citep{Mazzarella2012}.  
    Mrk 266 contains a confirmed dual AGN with $\sim\!6$ kpc separation and exhibits prominent, large-scale ionized winds \citep{Mazzarella2012}.  The southwest (SW) nucleus is classified as a Seyfert 2 and is Compton thick, while the northeast (NE) is a LINER and Compton thin \citep{Iwasawa2020}.  \fig{fig:MRK266-fcc} shows a false color image created from HST images with the blue channel as the F330W filter, green as the $\text{H}\alpha$ F673N filter, and red as the F814W filter.  The green channel is shown with a logarithmic scale to clearly show the extended $\text{H}\alpha$ features to the north and east.    
    \begin{figure}
	        % or pdf, eps, jpg if compiling using pdf latex
	        \includegraphics[width=\columnwidth]{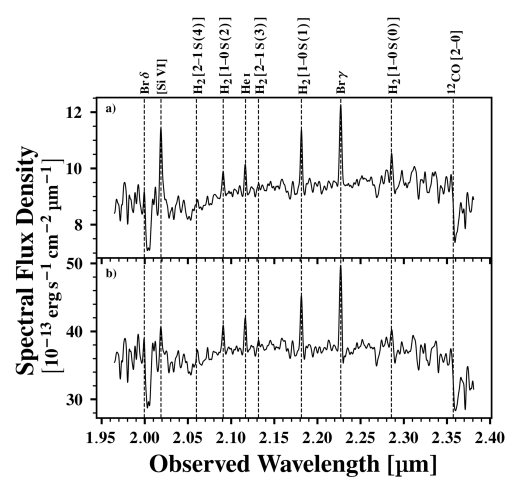}
            \caption{Integrated \textit{K}-band spectra of Mrk 266 SW with apertures of $0.5\arcsec$ (\textit{a}) and $1\arcsec$ (\textit{b}) in diameter centered on the AGN.  Wavelengths are not corrected for redshift.  The compact nature of the [\ion{Si}{vi}] emission line is evident since its flux does not increase significantly with larger apertures. }
            \label{fig:IntSpec_SW}
        \end{figure}
                \begin{figure}
                    % or pdf, eps, jpg if compiling using pdf latex
                    \includegraphics[width=\columnwidth]{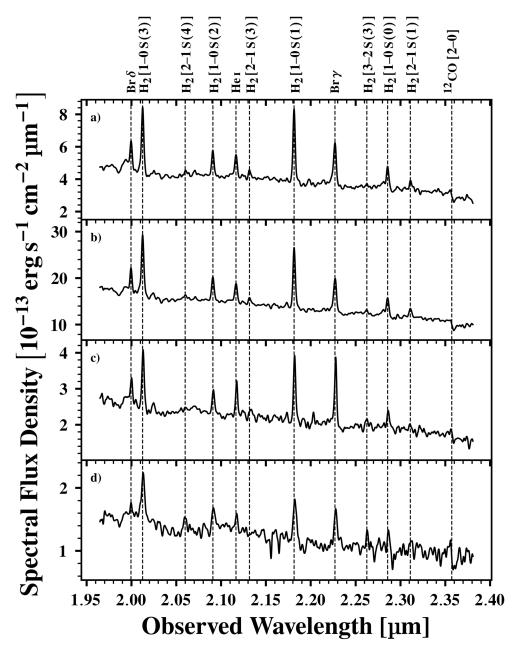}
                    \caption{Integrated \textit{K}-band spectra of Mrk 266 NE.  Wavelengths are not corrected for redshift.  Locations and apertures of the spectra are indicated in \fig{fig:IntSpec_NE_loc}.  (\textit{a}, blue) and (\textit{b}, red) are centered on the AGN with apertures of $0.5\arcsec$ and $1\arcsec$ respectively.  (\textit{c}, green) and (\textit{d}, magenta) both have apertures of $0.5\arcsec$.  The spectra show many $\text{H}_2$ emission lines, suggesting the presence of different mechanisms that make strong H$_2$ emission \citep{Black1987,Hollenbach1989,Maloney1996}.  In addition, the stellar absorption line $^{12}\text{CO}\,2\text{--}0$ is much weaker compared to the SW nucleus.}
                    \label{fig:IntSpec_NE}
                \end{figure}
                \begin{figure}
                    \centering
                    % or pdf, eps, jpg if compiling using pdf latex
                    \includegraphics[width=0.96\columnwidth]{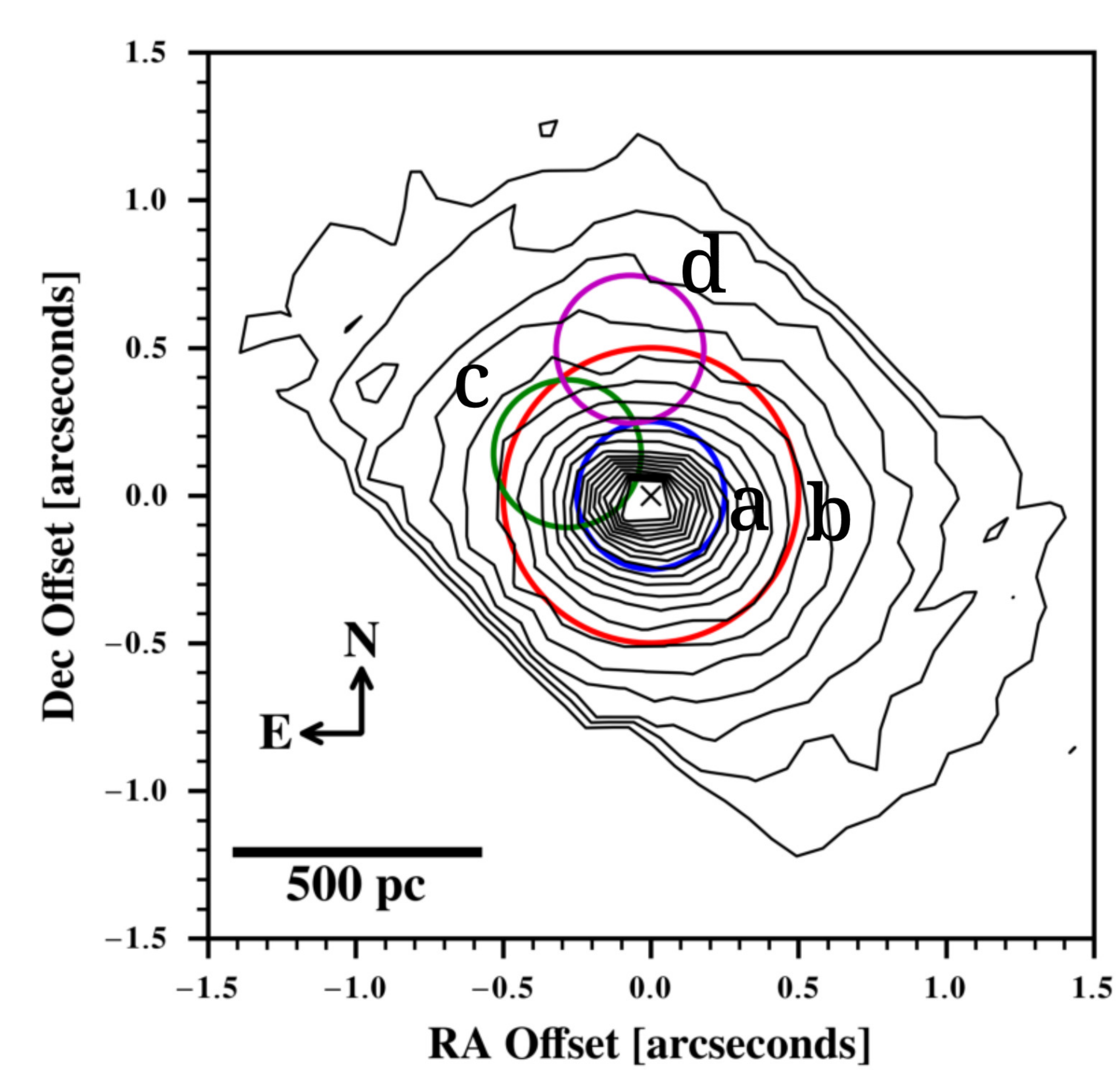}
                    \caption{Locations of spectra in \fig{fig:IntSpec_NE} with \textit{K}-band continuum contours of the NE nucleus. (\textit{a}) is indicated by the blue circle, (\textit{b}) red, (\textit{c}) green, and (\textit{d}) magenta.}
                    \label{fig:IntSpec_NE_loc}
                \end{figure}

Mrk 266 is an ideal galaxy for studying the relationships between mergers and AGN activity.  There are significant perturbations from the merger at \unit{\kilo\parsec} scales, but both nuclear regions remain distinct with different modes of AGN activity, given that the SW nucleus is $\sim\!30$ times more luminous than the NE in the X-ray ($2-10\,\text{kev}$) \citep{Iwasawa2020}.  The total molecular hydrogen gas mass for both nuclei is \qty{7.0e9}{\solarM} \citep{Imanishi2009,Mazzarella2012} indicating a large amount of gas available for AGN accretion.  In addition, the mass ratio of the two progenitor galaxies is $\sim\!1.25$, and according to simulations, a mass ratio of $\sim\!1$ increases the likelihood for the formation of dual AGN \citep{Steinborn2016,Volonteri2016}.  Both galaxies have a mass within the range of $5$\nobreakdash--\qty{6e10}{\solarM} \citep{Mazzarella2012}.   The relatively low redshift of $z=0.028$ allows for a high spatial resolution in which an angular resolution of $0.3\arcsec$ corresponds to a spatial resolution of approximately \qty{180}{\parsec}.  While Mrk 266 has been studied extensively using imaging and X-ray observations \citep{Leipski2006,Brassington2007,Imanishi2009,Mazzarella2012,Torres-Alba2018,Iwasawa2020}, we present here the first two-dimensional (2D) kinematic study of the gas and stars in the two nuclei using adaptive optics (AO)-assisted integral-field unit (IFU) spectroscopy, obtained with the OH-Suppressing Infra-Red Imaging Spectrograph (OSIRIS) at the Keck Observatory. The aim of this work is to identify and characterize for the first time the physical processes of feeding and feedback in the two nuclei of Mrk 266. A detailed characterization of the processes occurring in dual AGN is crucial to (i) understand the role of galaxy mergers in shaping the growth of bulges and SMBHs, and (ii) obtain critical clues about the conditions under which both SMBHs can be activated during these encounters to provide necessary guidance to theoretical models and future searches of dual and binary AGN.

This paper is organized as follows: In Section \ref{two}, the near-IR OSIRIS spectroscopic observations and the reduction process are described. In Section \ref{three}, the spatial distribution and 2D kinematics of the gas and stars in the two nuclei are presented. In Section \ref{Discussion} we discuss the interpretation of the kinematics observed in the two nuclei including kinematic modeling of the stars and gas. The summary and conclusions are finally discussed in Section \ref{five}. Throughout the paper we adopt the following parameters: a Hubble distance $D=120$ Mpc and scale of $582$ pc/arcsec, at the source redshift of $z=0.028$, with cosmological parameters corresponding to a $\Lambda$CDM universe with $\Omega_m=0.3$, $\Omega_\lambda=0.7$, and H$_0=70.3$ km s$^{-1}$ Mpc$^{-1}$. 

\section{Observations and Data Reduction} \label{two}
\subsection{Observations}
    The NE and SW nuclei of Mrk 266 were observed with the Keck OSIRIS integral field unit (IFU) spectrograph on 9 February 2015 and 23 April 2016 with an airmass of $1.14$ and $1.24$, respectively.  For both nuclei, the broad \textit{K}-band filter was used, covering the range from \qty{1965}{\nano\meter} to \qty{2381}{\nano\meter} with $1665$ channels.  The \qty{0.1}{\arcsecond} per pixel plate scale was used, giving a field-of-view (FOV) of $\qty{1.6}{\arcsecond}\,\times\,\qty{6.4}{\arcsecond}$.  The integration time for each frame was \qty{600}{\second}, and $4$ frames were coadded, giving a total integration time of \qty{2400}{\second} for both nuclei.  Laser guide star (LGS) adaptive optics were used for both observations, which resulted in an angular resolution at \qty{2.2}{\micro\meter} of approximately \qty{0.31}{\arcsecond} and \qty{0.20} {\arcsecond} FWHM for the NE and SW nuclei respectively. These values were measured by fitting a two-component 1D Gaussian to the flux profile extracted along the major axis of the nuclei and taking the FWHM of the narrower component, thereby measuring the width of the unresolved AGN while removing contamination from the extended host galaxy emission. The spectral resolving power, $R$, was measured by fitting a 1D Gaussian profile to atmospheric OH emission lines \citep{OH_Lines}.  An $R$ of $3250$ was measured for the SW nucleus and $2680$ for the NE nucleus, corresponding to velocity resolutions at \qty{2.2}{\micro\meter} of \qty{92}{\kilo\meter\per\second} and \qty{112}{\kilo\meter\per\second} FWHM for the SW and NE nuclei, respectively.  This corresponds to minimum measurable velocity dispersions of \qty{39}{\kilo\meter\per\second} for the SW and \qty{48}{\kilo\meter\per\second} for the NE.

        \begin{figure*}
	        % or pdf, eps, jpg if compiling using pdf latex
	        \includegraphics[width=\textwidth]{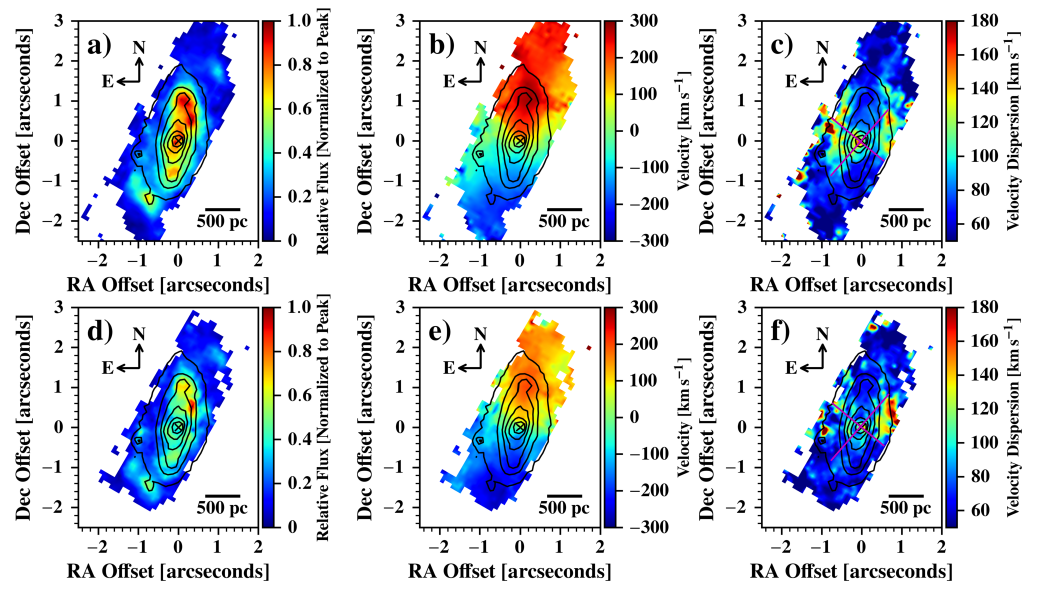}
            \caption{Flux (\textit{a} and \textit{d}), velocity (\textit{b} and \textit{e}), and velocity dispersion (\textit{c} and \textit{f}) maps of $\text{Br}\gamma$ (\textit{top}) and \ion{He}{i} (\textit{bottom}) for the SW nucleus.  Contours of the \textit{K}-band continuum were measured via the mean of each spectrum and range linearly from $0.5\text{\nobreakdash{--}}\num{7.5e-14}\,\text{erg}\,\unit{\second^{-1}\centi\meter^{-2}\micro\meter^{-1}}$, and the '$\times$' indicates the location of the AGN. In both cases, most of the emission is concentrated in a ring approximately \qty{600}{\parsec} from the AGN.  The velocity maps are dominated by rotation, and the dispersion maps show a biconical structure (indicated by the magenta lines) oriented perpendicular to the rotation and centered on the AGN, suggesting the presence of an AGN-driven outflow (see Section \ref{outflow}).  Bilinear interpolation was used for these and all subsequent maps.}
            \label{fig:BrGammaSW}
        \end{figure*}
        
        \begin{figure*}
	        % or pdf, eps, jpg if compiling using pdf latex
	        \includegraphics[width=\textwidth]{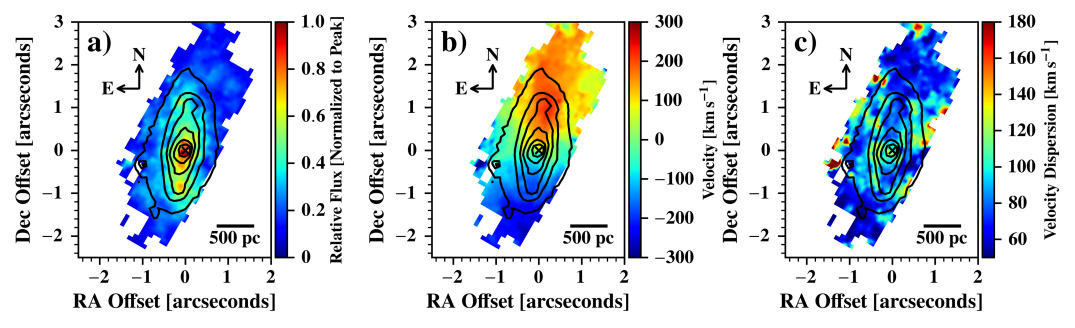}
            \caption{Flux (\textit{a}), velocity (\textit{b}), and velocity dispersion (\textit{c}) maps of $\text{H}_2$ for the SW nucleus.  Contours of the \textit{K}-band continuum range linearly from $0.5\text{\nobreakdash{--}}\num{7.5e-14}\,\text{erg}\,\unit{\second^{-1}\centi\meter^{-2}\micro\meter^{-1}}$, and the '$\times$' indicates the location of the AGN.  The velocity map is dominated by rotation, suggesting that the molecular gas has not decoupled from the rotation of the stars.}
            \label{fig:H2SW}
        \end{figure*}

        \begin{figure*}
	        % or pdf, eps, jpg if compiling using pdf latex
	        \includegraphics[width=\textwidth]{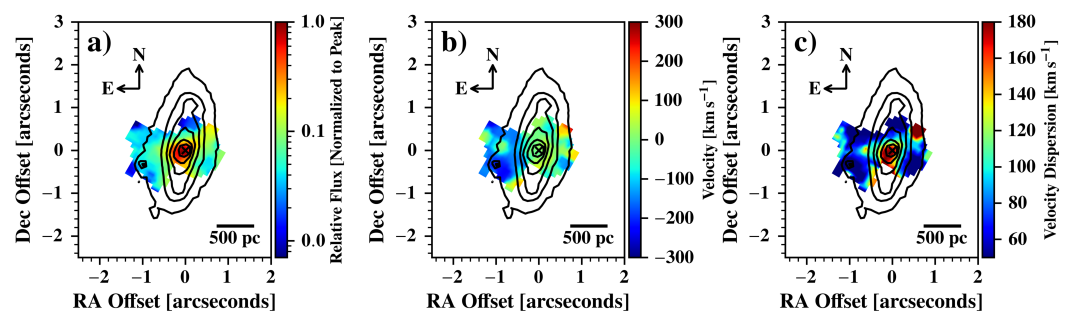}
            \caption{Flux (\textit{a}), velocity (\textit{b}), and velocity dispersion (\textit{c}) maps of [\ion{Si}{vi}] for the SW nucleus. Contours of the \textit{K}-band continuum range linearly from $0.5\text{\nobreakdash{--}}\num{7.5e-14}\,\text{erg}\,\unit{\second^{-1}\centi\meter^{-2}\micro\meter^{-1}}$, and the '$\times$' indicates the location of the AGN. The flux map is plotted with a log color scale to emphasize its biconical structure.  Since the [\ion{Si}{vi}] emission is centered on the AGN, has a biconical structure, perpendicular to the rotation of $\text{H}_2$ and $\text{Br}\gamma$, and exhibits a velocity gradient that is also perpendicular to the rotation, the emission is very likely an AGN-driven outflow (see Section \ref{outflow}).  Due to its compact size, the outflow is likely very young, indicating that the AGN was triggered during the merger event.}
            \label{fig:SiVISW}
        \end{figure*}

\subsection{Data Reduction and Emission Line Fitting}
    The data were reduced using the Keck OSIRIS Data Reduction Pipeline \citep{OSIRIS_DRP_1,OSIRIS_DRP_2}.  First, the subtract frame routine was used to subtract a dark frame from each of the science frames.  Then, the adjust channel levels, remove crosstalk, glitch identification, and clean cosmic rays routines were all applied to the raw data.  Next, the extract spectra, assemble data cube, and correct dispersion routines were applied to extract the data cube.  Finally, the scaled sky subtraction routine was used to remove OH lines, and the divide by star spectrum routine corrected for atmospheric absorption.  The star spectrum was created by manually removing emission lines from a spectrum extracted from a calibration star and dividing by a black-body curve.

    After reducing the data cubes, flux calibrations were performed using  observations of the stars HD 87035 for the SW nucleus and HD 120405 for the NE nucleus.  The SIMBAD astronomical database \citep{SIMBAD} was used to retrieve absolute \textit{K}-band magnitudes of \num{7.523(20)} for HD 87035 and \num{6.914(16)} for HD 120405 \citep{2MASS}.  The Vega \textit{K}-band zeropoint spectral flux density of \qty{3.961e-11}{\erg\per\second\per\centi\meter\squared\per\angstrom} \citep{Bessell1998} and the bandwidth of $0.41\ \micron$ were used to determine the sensitivity function for the OSIRIS observations.

    To create the kinematic maps for each emission line, AstroPy \citep{astropy:2013,astropy:2018,astropy:2022} was used to fit a Gaussian plus continuum model to each emission line after applying a constant $3\times3\times3$ smoothing kernel.  The continuum was modeled as either a first or second degree polynomial, depending on the local behavior.  The Levenberg-Marquardt least squares non-linear fitting algorithm was used to fit the model.  Errors for each parameter of the Gaussian component were estimated with the standard deviations calculated from the covariance matrix returned by the fitting routine.  Pixels were removed if the fitted parameters were at the bounds, if the flux of the Gaussian profile was lower than the standard deviation of the residuals, or if the fitting routine returned an error indicating a failure to reduce squared residuals after multiple iterations.  Flux maps were generated by integrating the fitted gaussian for each pixel.  However, the flux values given in Table~\ref{tab:EmissionLineFlux} were measured by directly integrating the emission line, and errors were measured using the standard deviation of the residuals after subtracting a Gaussian plus continuum model and then propagating the standard deviation through the flux calculations.  For the velocity measurements, the zero point was set at the velocity of the AGN by manually measuring the center of each emission line in an integrated $0.5\arcsec$ aperture centered on the AGN.

\section{Results} \label{three}

    \subsection{K-band Spectra of the Two Nuclei}
        \fig{fig:IntSpec_SW} shows integrated spectra with apertures of $0.5\arcsec$ and $1\arcsec$ centered on the AGN in Mrk 266 SW.  All detected emission lines and fluxes for multiple apertures are shown in Table~\ref{tab:EmissionLineFlux}.  The integrated \textit{K}-band Vega magnitude is $12.5$ for a circular aperture $1\arcsec$ in diameter centered on the AGN.  For comparison, the 2MASS $K_s$-band magnitude for a $3\arcsec\times3\arcsec$ FOV is 12.7.  We report the detection of the \qty{167}{\eV} coronal line $[\ion{Si}{vi}]$ with a flux of \num{1.18(26)e-15} and \qty{1.24(11)e-15}{erg\,s^{-1}\,cm^{-2}} in $1\arcsec$ and $0.5\arcsec$ diameter apertures centered on the AGN.  The fluxes for the two apertures are statistically equivalent, indicating that the $[\ion{Si}{vi}]$ emission is concentrated near the AGN.

        In addition, we detect three $\text{H}_2$ lines, \ion{He}{i}, $\text{Br}\gamma$, and the stellar $^{12}\text{CO}\,[2\text{--}0]$ absorption band head.  Flux, velocity, and dispersion maps for $[\ion{Si}{vi}]$, \ion{He}{i}, $\text{Br}\gamma$, and $\text{H}_2$ are presented in the following sections.  Extraction of stellar kinematics typically requires multiple absorption band heads, but the $^{12}\text{CO}\,[3\text{--}1]$ and $^{12}\text{CO}\,[4\text{--}2]$ band heads at \qty{2322.7}{\nano\meter} and \qty{2352.5}{\nano\meter} in the rest frame are redshifted beyond the range of OSIRIS. However, the strong signal-to-noise ratio (SNR) of $^{12}\text{CO}\,[2\text{--}0]$ in the SW nucleus allowed extraction of the stellar kinematics presented below.

        \begin{figure}
	        % or pdf, eps, jpg if compiling using pdf latex
            \centering
	        \includegraphics[width=\columnwidth]{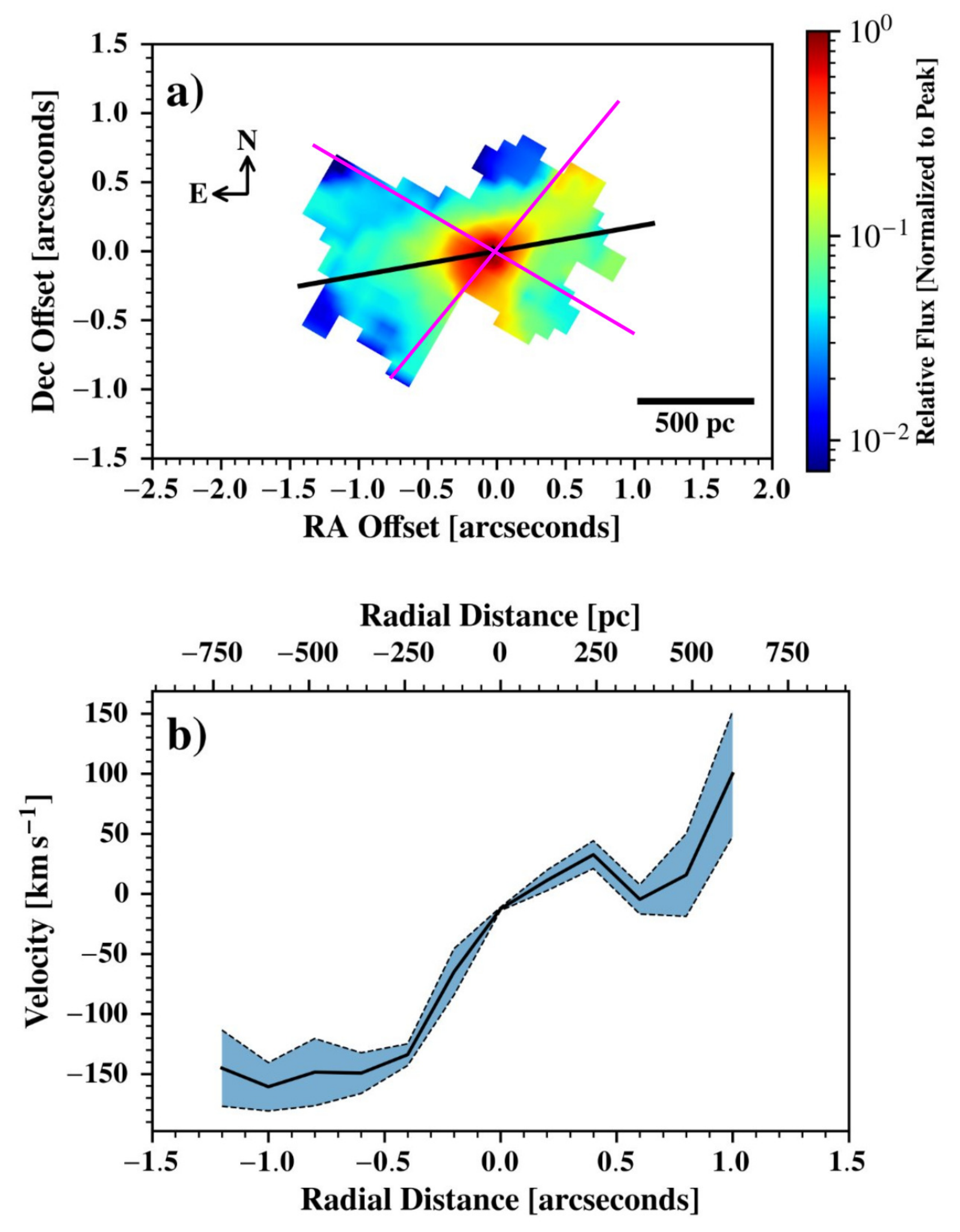}
            \caption{$[\ion{Si}{vi}]$ outflow velocity profile for the SW nucleus.  (\textit{a}) Flux map of $[\ion{Si}{vi}]$ with log color scale and a line overlaid to indicate the position angle of the outflow ($100$ degrees).  The magenta lines match those in Fig.~\ref{fig:BrGammaSW} and correspond to an opening angle of \qty{80}{\degree}.  (\textit{b}) Radial velocity profile centered at the peak of $[\ion{Si}{vi}]$ emission.  The profile was generated by taking the median of the velocities within half of a $0.2\arcsec$ wide annulus.  Pixels east of the peak of $[\ion{Si}{vi}]$ emission were given a negative radial distance, and pixels west of the peak were given a positive radial distance.  The shaded region indicates uncertainty measured using the median absolute deviation.}  
            \label{fig:SiVISW_velProf}
        \end{figure}

        \fig{fig:IntSpec_NE} shows integrated spectra of the NE nucleus with locations and apertures indicated by \fig{fig:IntSpec_NE_loc}.  The \textit{K}-band Vega magnitude is $13.6$ for an aperture $1\arcsec$ in diameter, and the 2MASS $K_s$-band magnitude in a $3\arcsec\times3\arcsec$ FOV is $13.0$.  We detect eight $\text{H}_2$ lines, \ion{He}{i}, both $\text{Br}\gamma$ and $\text{Br}\delta$, and the $^{12}\text{CO}\,[2\text{--}0]$ bandhead.  The $\text{Br}\gamma$ line has a slight blue tail in $(c)$ and a slight red tail in $(d)$, but for consistency, a single Gaussian component was used to fit all lines.  Unlike the SW nucleus, there are no high-ionization lines, and there are more transitions of $\text{H}_2$. This suggests that the AGN in the NE nucleus is in a weaker mode of accretion. In addition, the variety of $\text{H}_2$ lines suggests the presence of shocks \citep{Mazzarella2012} and other mechanisms that can generate strong H$_2$ rovibrational emission lines such as ultraviolet and collisional excitation,  \citep{Black1987,Hollenbach1989,Maloney1996}.  While the $^{12}\text{CO}\,[2\text{--}0]$ bandhead is present, the low SNR for small apertures prevented reliable measurement of stellar kinematics for the NE nucleus.

    \subsection{Flux, Velocity, and Dispersion Maps for the SW Nucleus}
        Figures $5-7$ shows emission line maps of Mrk 266 SW for $\text{H}_2$, $\text{Br}\gamma$, $[\ion{Si}{vi}]$, and \ion{He}{i}.  In all maps, the '$\times$' indicates the position of the AGN inferred from the peak of \textit{K}-band continuum emission, and the $x$ and $y$ axes indicate the offset from the AGN in arcseconds.  All flux maps are normalized to peak, and all maps are oriented with north up and east left.  No secondary smoothing was applied to the $\text{Br}\gamma$ or $\text{H}_2$ maps besides bilinear pixel interpolation.  A $3\times3$ uniform smoothing kernel was applied to the $[\ion{Si}{vi}]$ maps due to a low SNR outside of the nuclear region.  Pixels for which the fitting routine was unsuccessful or that were likely to fit noise have been left white.  For comparison, all maps are plotted with \textit{K}-band continuum contours measured via the mean of each spectrum unless otherwise indicated.
    \subsubsection{Low Ionization Gas}

        \fig{fig:BrGammaSW} shows the maps for $\text{Br}\gamma$ and \ion{He}{i} for the SW nucleus.  In the $\text{Br}\gamma$ flux map, we see several clumps that are likely star forming regions, with the peak of emission $1\arcsec$ north of the AGN.  The velocity map shows that the dominant kinematic component is ordered rotation with apparently little perturbation from the galaxy merger.  The maximum and minimum velocities are \qty{289}{\kilo\meter\per\second} and \qty{-326}{\kilo\meter\per\second}, and the kinematic minor axis passes through the AGN.  The velocity dispersion map is mostly uniform with values ranging from $60-80$ \unit{\kilo\meter\per\second}, which are consistent with a geometrically thin disk. In addition, the dispersion map clearly shows a second kinematic component with a biconical shape, centered on the AGN and aligned perpendicular to the rotation of the galaxy.  The maximum velocity dispersion within the biconical component is \qty{164}{\kilo\meter\per\second} with an average of \qty{118}{\kilo\meter\per\second}, and the opening angle is \qty{84} degrees with position angle \qty{100} degrees east of north.

        The \ion{He}{i} emission closely follows that of $\text{Br}\gamma$, with the peak flux approximately $1\arcsec$ northwest of the AGN.  The rotation is slower, however, matching that of the $\text{H}_2\,[1\text{--}0\,S(1)]$ described below with maximum and minimum velocities of \qty{202}{\kilo\meter\per\second} and \qty{-279}{\kilo\meter\per\second}.  The velocity dispersion map is mostly uniform with values ranging from $60-80$ km s$^{-1}$ with an additional component west of the AGN aligned with the second component in the $\text{Br}\gamma$ velocity dispersion map with a maximum dispersion of \qty{181}{\kilo\meter\per\second}.
        
        \begin{figure}
            \centering
            \includegraphics[width=\columnwidth]{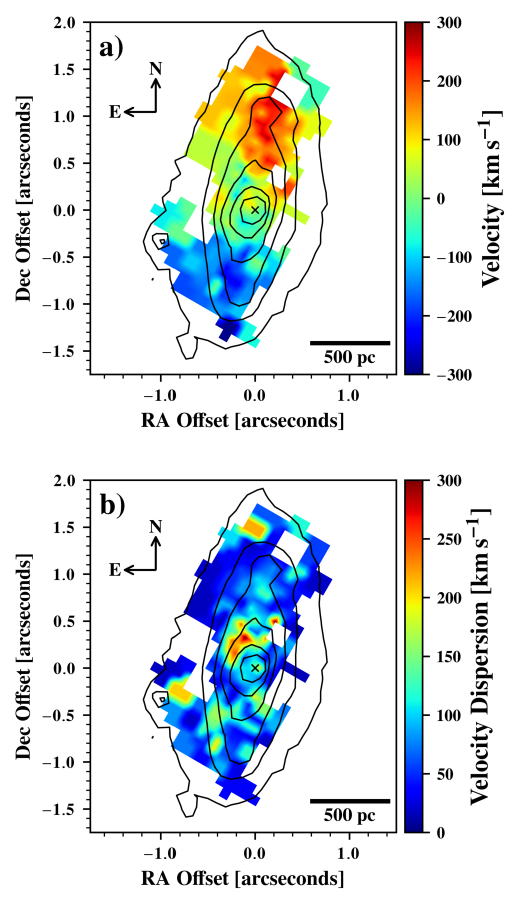}
            \caption{Velocity (\textit{a}) and velocity dispersion (\textit{b}) maps for the stellar content in the SW nucleus.  Prior to fitting the $^{12}\text{CO}\,2\text{--}0$ stellar absorption line with pPXF \citep{Cappellari2004,Cappellari2017}, Voronoi binning was applied to ensure a minimum SNR of 25 for the \textit{K}-band continuum flux. Contours of the \textit{K}-band continuum range linearly from $0.5\text{\nobreakdash{--}}\num{7.5e-14}\,\text{erg}\,\unit{\second^{-1}\centi\meter^{-2}\micro\meter^{-1}}$, and the '$\times$' indicates the location of the AGN.}
            \label{fig:StellVelSW}
        \end{figure}       
    \subsubsection{Molecular Gas}
        \fig{fig:H2SW} shows the flux, velocity, and velocity dispersion maps of the $\text{H}_2\,[1\text{--}0\,S(1)]$ emission line.  The peak of $\text{H}_2$ emission is found near the AGN, and the flux map shows an extended structure aligned with the rotation of the galaxy.  The $\text{H}_2$ velocity map has a similar structure to the $\text{Br}\gamma$ velocity map in \fig{fig:BrGammaSW}, with a clear lack of strong perturbations caused by the merger.  The maximum and minimum velocities are \qty{221}{\kilo\meter\per\second} and \qty{-268}{\kilo\meter\per\second}, slightly less than the maximum rotation velocities of $\text{Br}\gamma$.  The $\text{H}_2$ velocity dispersion map does not clearly show the biconical structure found in the $\text{Br}\gamma$ dispersion map, and the dispersion is slightly higher in $\text{H}_2$ with a maximum value of \qty{176}{\kilo\meter\per\second} near the AGN and an average of \qty{103}{\kilo\meter\per\second}.

        \begin{figure*}
            \includegraphics[width=\textwidth]{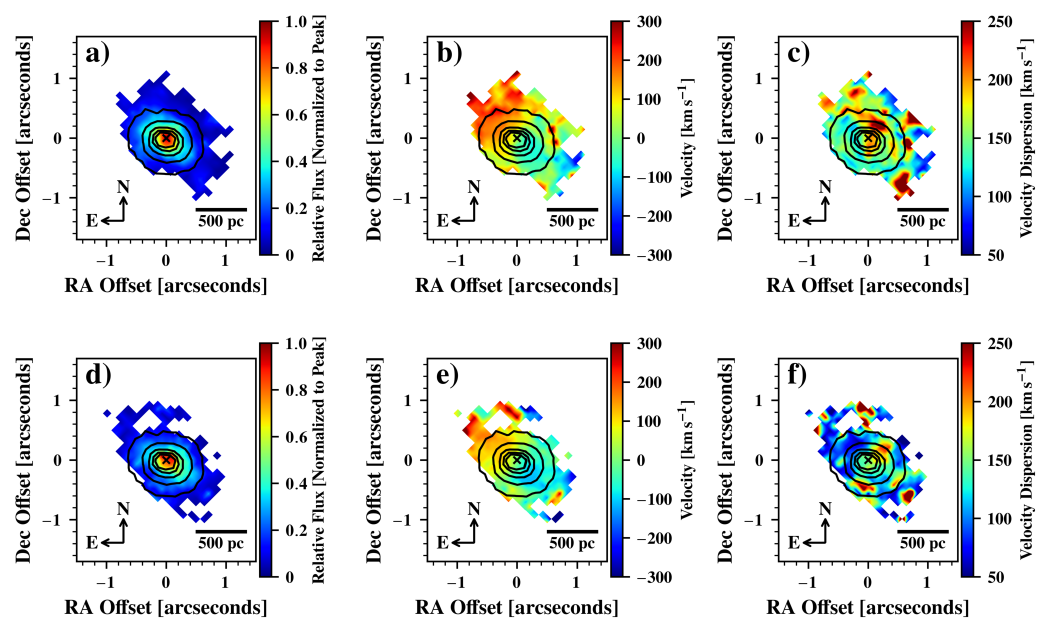}
            \caption{Flux (\textit{a} and \textit{d}), velocity (\textit{b} and \textit{e}), and velocity dispersion (\textit{c} and \textit{f}) maps of $\text{Br}\gamma$ (\textit{top}) and \ion{He}{i} (\textit{bottom}) for the NE nucleus.  Contours of the \textit{K}-band continuum were measured via the mean of each spectrum and range linearly from $1\text{\nobreakdash{--}}\num{3e-14}\,\text{erg}\,\unit{\second^{-1}\centi\meter^{-2}\micro\meter^{-1}}$, and the '$\times$' indicates the location of the AGN.  The flux for both $\text{Br}\gamma$ and \ion{He}{i} is concentrated near the AGN, and the velocity maps primarily show rotation.  }
            \label{fig:BrGammaNE}
        \end{figure*}
        \begin{figure*}
            % or pdf, eps, jpg if compiling using pdf latex
            \includegraphics[width=\textwidth]{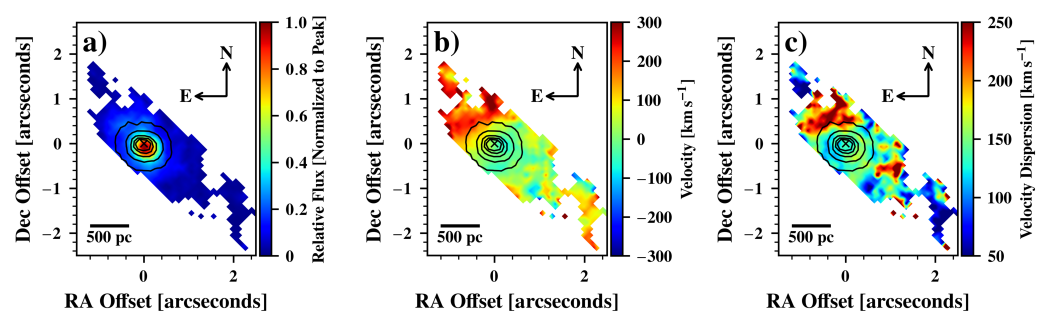}
            \caption{Flux (\textit{a}), velocity (\textit{b}), and velocity dispersion (\textit{c}) maps of $\text{H}_2$ for the NE nucleus.  Contours of the \textit{K}-band continuum range linearly from $1\text{\nobreakdash{--}}\num{3e-14}\,\text{erg}\,\unit{\second^{-1}\centi\meter^{-2}\micro\meter^{-1}}$, and the '$\times$' indicates the location of the AGN.  The molecular gas has an extended structure with components that are decoupled from the rotation of the galaxy due to merger-related perturbations.}
            \label{fig:H2NE}
        \end{figure*}

    \subsubsection{High Ionization Gas}
        \fig{fig:SiVISW} shows $[\ion{Si}{vi}]$ maps for the SW nucleus, and \fig{fig:SiVISW_velProf} shows extracted velocity profiles.  All 3 maps have been smoothed with a $3\times3$ uniform kernel before pixel interpolation was applied.  The ionization potential of $[\ion{Si}{vi}]$ is \qty{167}{\electronvolt}, indicating the existence of soft X\nobreakdash-rays \citep{Veilleux1997}. The flux map shows a very strong concentration of $[\ion{Si}{vi}]$ emission near the AGN and an extended structure perpendicular to the major axis of the galaxy.  The projected length of this structure is $1.3\arcsec$ from the AGN, corresponding to \qty{0.8(1)}{\kilo\parsec}, with position angle \qty{100}{\degree} and opening angle \qty{80}{\degree}, identical to the biconical structure in $\text{Br}\gamma$ velocity dispersion (\fig{fig:BrGammaSW} \textit{c}).  The velocity map shows a gradient that is perpendicular to the rotation of the galaxy with a maximum radial velocity along the center black line in \fig{fig:SiVISW_velProf} of \qty{150}{\kilo\meter\per\second}.  The maximum velocity dispersion is \qty{225}{\kilo\meter\per\second} with an average of \qty{95}{\kilo\meter\per\second}.  An inclination angle of \qty{70} degrees was measured for Mrk 266 with kinematic modelling of the stellar kinematics (see Section \ref{stars}). Assuming the [SiVI] velocities are perpendicular to the plane of the galaxy, the maximum velocity in this region is \qty{400}{\kilo\meter\per\second}, and the distance from the AGN to the eastern edge of the biconical structure is \qty{0.85}{\kilo\parsec}.

    \subsubsection{Stars}      \label{stars}

        Stellar kinematics for the SW nucleus were fit using the software pPXF developed by \cite{Cappellari2004} and \cite{Cappellari2017}.  Stellar templates where collected from the Gemini GNIRS library of \textit{K}-band spectra, and prior to fitting the templates, Voronoi binning was applied to ensure a minimum SNR of 25 for the \textit{K}-band continuum flux.  \fig{fig:StellVelSW} shows the results.  The velocity map aligns closely to that of $\text{H}_2$ in \fig{fig:H2SW}, with the primary kinematic component being rotation.  The velocity dispersion has a maximum value of \qty{310}{\kilo\meter\per\second} approximately \qty{500}{\parsec} northeast of the AGN.  The average velocity dispersion in a circular region centered on the AGN with a diameter of \qty{1}{\kilo\parsec} is \qty{120(10)}{\kilo\meter\per\second}.

        The software pPXF \citep{Cappellari2004,Cappellari2017} fits a model comprising of a weighted sum of stellar templates convolved with a Gauss-Hermite series with multiplicative and additive orthogonal polynomials.  The fitting procedure utilizes a penalized maximum likelihood estimator such that non-Gaussian solutions are penalized.  Errors for the fit are estimated by pPXF using the covariance matrix of the standard errors assuming it is diagonal when minimized.  Bad pixels were manually thrown out if the fitted velocity or velocity dispersion were at the bounds or if the SNR of the continuum was below 0.5, thereby removing large bins at the edges of the field of view created by the Voronoi binning.

    \subsection{Flux, Velocity, and Dispersion Maps for the NE Nucleus}
        
        \subsubsection{Low Ionization Gas}
            
              \begin{table*}
                            \centering
                            \renewcommand{\arraystretch}{1.2}
                            \caption{Kinematic modelling results for the SW and NE nuclei.  The model was that for an exponential rotating disk described in \citet{MocKinG} with center fixed at the AGN.  PA and inclination were measured using contours of the HST F110W image, and these results were used to constrain the model when fitting the stars (SW) or the molecular gas (NE).  The results from the stars or molecular gas were used to fix the PA and inclination for all other models.  $1\sigma$ errors are reported.}
                            \label{tab:MocKinG-Results}
                            \begin{tabular}{cccc} % four columns, alignment for each
                                \hline
                                        \multirow{2}{*}{Parameter} & \multirow{2}{*}{Stars*} & \multirow{2}{*}{Molecular Gas} & \multirow{2}{*}{Low-ionization Gas} \\
                                & &  &  \\
                                \hline
                                \multicolumn{4}{c}{SW Nucleus} \\
                                \hline
                                Morphological PA\textsuperscript{\textdagger}  &   \qty{-7(5)}{\degree} & \qty{-7(5)}{\degree}  & \qty{-7(5)}{\degree}  \\
                                Morphological Inclination\textsuperscript{\textdagger}    & \qty{66(5)}{\degree}& \qty{66(5)}{\degree} & \qty{66(5)}{\degree} \\
                                Kinematic PA   & \qty{-12.8(2)}{\degree}& \qty{-12.8(2)}{\degree} & \qty{-12.8(2)}{\degree} \\
                                Kinematic Inclination& \qty{72.2(4)}{\degree}& \qty{72.2(4)}{\degree} & \qty{72.2(4)}{\degree}\\
                                Radius of Max Velocity & \qty{0.99(15)}{\kilo\parsec} & \qty{1.21(15)}{\kilo\parsec} & \qty{0.95(15)}{\kilo\parsec} \\
                                Max Velocity & \qty{209(2)}{\kilo\meter\per\second} & \qty{224.6(4)}{\kilo\meter\per\second} & \qty{269.4(3)}{\kilo\meter\per\second} \\
                                Velocity Offset & \qty{-9.1(6)}{\kilo\meter\per\second} & \qty{-28.4(2)}{\kilo\meter\per\second} & \qty{30.9(2)}{\kilo\meter\per\second} \\
                                \hline
                                \multicolumn{4}{c}{NE Nucleus} \\
                                \hline
                                Morphological PA\textsuperscript{\textdagger}  & \textemdash & \qty{53(2)}{\degree} & \qty{53(2)}{\degree} \\
                                Morphological Inclination\textsuperscript{\textdagger}  & \textemdash & \qty{36(5)}{\degree} & \qty{36(5)}{\degree} \\
                                Kinematic PA  & \textemdash & \qty{43.5(7)}{\degree} & \qty{43.5(7)}{\degree} \\
                                Kinematic Inclination  & \textemdash & \qty{33.6(4)}{\degree} & \qty{33.6(4)}{\degree} \\
                                Radius of Max Velocity & \textemdash & \qty{0.27(15)}{\kilo\parsec} & \qty{0.30(15)}{\kilo\parsec} \\
                                Max Velocity & \textemdash & \qty{201(3)}{\kilo\meter\per\second} & \qty{270(4)}{\kilo\meter\per\second} \\
                                Velocity Offset & \textemdash & \qty{54.8(6)}{\kilo\meter\per\second} & \qty{75.6(6)}{\kilo\meter\per\second} \\
                                \hline
                                \multicolumn{4}{p{10.5cm}}{*Stellar kinematics were not extracted for the NE nucleus due to the low SNR of the $^{12}\text{CO}\,[2\text{--}0]$ absorption bandhead and lack of additional stellar absorption lines redshifted beyond the spectral range of Keck/OSIRIS.} \\
                                \multicolumn{4}{p{10.5cm}}{\textsuperscript{\textdagger}Morphological PA and inclination measured by fitting an ellipse to contours of HST F110W.}
                            \end{tabular}
                        \end{table*}
            \begin{figure*}
                        \includegraphics[width=\textwidth]{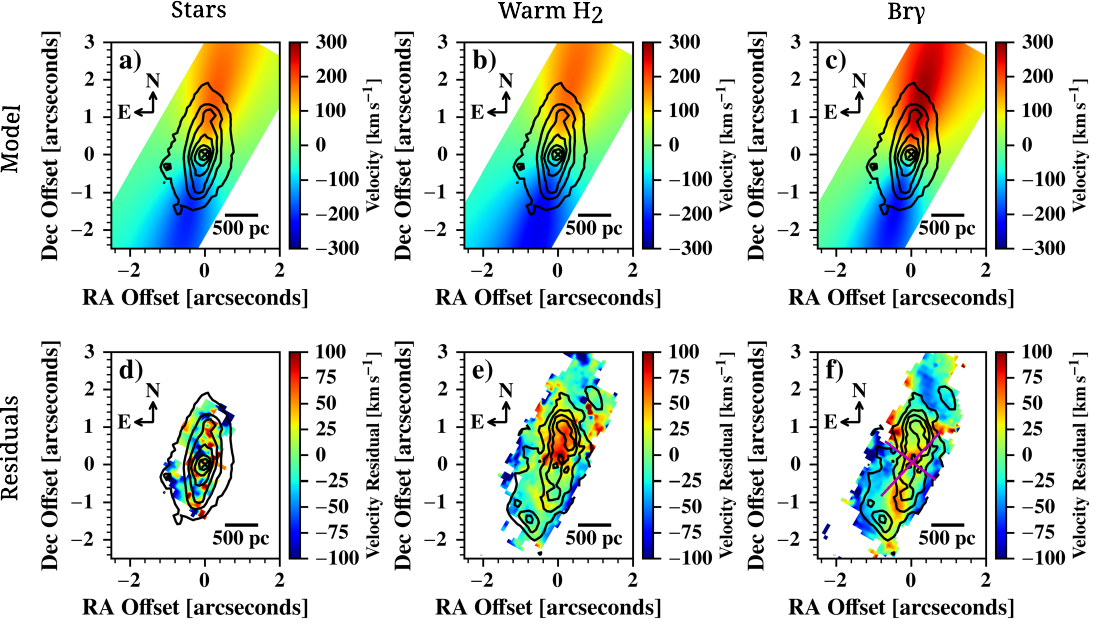}
                        \caption{Kinematic modelling results for the SW nucleus.  Contours for (\textit{a})\nobreakdash{--}(\textit{d}) are \textit{K}-band continuum ranging linearly from $0.5\text{\nobreakdash{--}}\num{7.5e-14}\,\text{erg}\,\unit{\second^{-1}\centi\meter^{-2}\micro\meter^{-1}}$ while those for (\textit{e}) and (\textit{f}) are $\text{Br}\gamma$ relative flux ranging linearly from $25\%$ to $80\%$ of the maximum. \textit{Top:} Velocity model for stars (\textit{a}), $\text{H}_2$ (\textit{b}), and $\text{Br}\gamma$ (\textit{c}).  \textit{Bottom:} Velocity residuals for stars (\textit{d}), $\text{H}_2$ (\textit{e}), and $\text{Br}\gamma$ (\textit{f}).  The redshifted velocity residuals for $\text{H}_2$ suggest an inflow of gas from a circumnuclear ring of star formation.  In the $\text{Br}\gamma$ velocity residuals, the blueshifted region east of the AGN is likely related to the outflow, and the redshifted region south of the AGN is associated with stellar winds.  The magenta lines in (\textit{f}) indicate the position of the [\ion{Si}{vi}] outflow.}
                        \label{fig:VelModel}
                    \end{figure*}

            \fig{fig:BrGammaNE} shows $\text{Br}\gamma$ and \ion{He}{i} flux, velocity, and dispersion maps for the NE nucleus.  The emission is confined to approximately \qty{500}{\parsec} from the AGN.  The velocity maps primarily show rotation, and there is no clear structure in the dispersion maps.  \ion{He}{i} has an RMS velocity of \qty{152}{\kilo\meter\per\second}, and  $\text{Br}\gamma$ has an RMS velocity of \qty{200}{\kilo\meter\per\second}.  The mean velocity dispersion is \qty{127}{\kilo\meter\per\second} and \qty{158}{\kilo\meter\per\second} for \ion{He}{i} and $\text{Br}\gamma$ respectively, and the maximum velocity dispersion is \qty{428}{\kilo\meter\per\second} and \qty{415}{\kilo\meter\per\second}.

            \subsubsection{Molecular Gas}

                \fig{fig:H2NE} shows the $\text{H}_2\,[1\text{--}0\,S(1)]$ line for the NE nucleus with strong evidence for non-circular motions. Towards the southwest, there is a redshifted region that is inconsistent with the rotation of the host galaxy.  To the northeast of the AGN, there is a region with high velocity dispersion, approximately \qty{255}{\kilo\meter\per\second}, and the average velocity dispersion is \qty{143}{\kilo\meter\per\second}.  The maximum velocity is \qty{282}{\kilo\meter\per\second} \qty{0.9}{\arcsecond} north of the AGN, and the minimum is \qty{-72}{\kilo\meter\per\second} \qty{0.2}{\arcsecond} west of the AGN.

        \begin{figure*}
            \includegraphics[width=\textwidth]{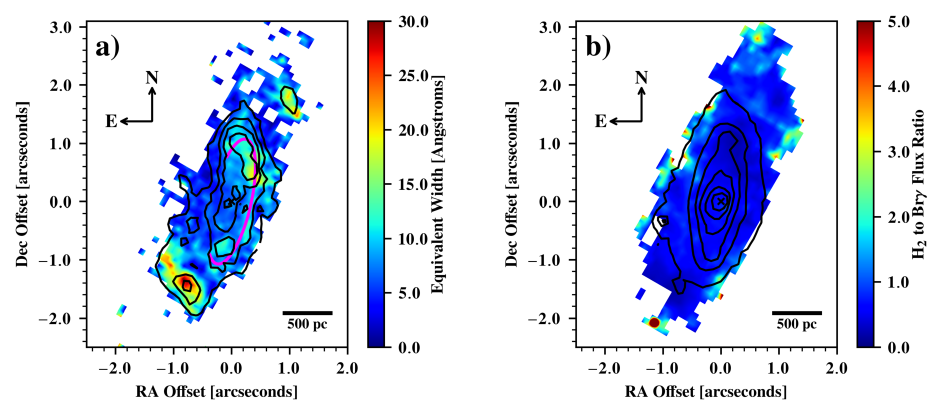}
            \caption{Equivalent width of $\text{Br}\gamma$ (\textit{a}) and $\text{H}_2$-to-$\text{Br}\gamma$ flux ratio (\textit{b}) for the SW nucleus.  Contours for (\textit{a}) are $\text{Br}\gamma$ relative flux ranging linearly from $25\%$ to $80\%$ of the maximum while contours for (\textit{b}) are \textit{K}-band continuum ranging linearly from $0.5\text{\nobreakdash{--}}\num{7.5e-14}\,\text{erg}\,\unit{\second^{-1}\centi\meter^{-2}\micro\meter^{-1}}$.  Equivalent width of $\text{Br}\gamma$ can be used to trace regions of star formation \citep{Pasha2020}.  Thus, the magenta ellipse likely traces a ring of star formation centered on the AGN with inclination and position angle matching that measured from the kinematic modelling.  The radius is approximately \qty{600}{\parsec}.  The low, flat $\text{H}_2$-to-$\text{Br}\gamma$ flux ratio map shows no evidence of shocked molecular gas in the SW nucleus.}
            \label{fig:H2BrGammaRatioSW}
        \end{figure*}
    \begin{figure*}
        \includegraphics[width=0.85\textwidth]{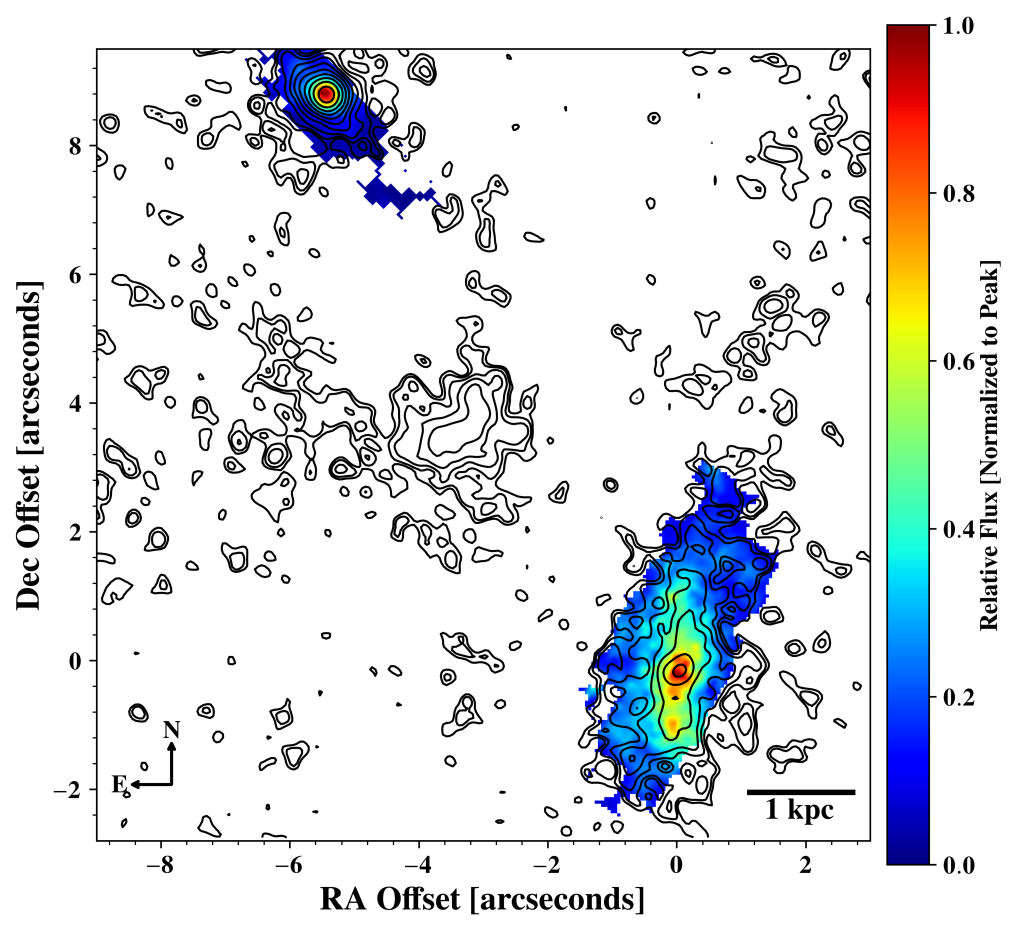}
        \caption{$\text{H}_2$ flux with VLA \qty{3.6}{\centi\meter} radio continuum contours obtained from \citet{Condon1991} to show connection to large scale structure.  Contour lines are plotted with log scale and range from $0.7\%$ to $70\%$ of the maximum.  The minimum contours correspond to $\sim\!\!1\sigma$ above the background.  RA and Dec are relative to SW AGN, and the $\text{H}_2$ flux for each nucleus is normalized independently.  The region of radio continuum emission between the two nuclei is likely the interface of the two galaxies \citep{Mazzarella2012}.  In the SW, the radio continuum emission is aligned with the disk of the galaxy with no evidence of an AGN-driven outflow extending beyond the FOV of Keck/OSIRIS.}
        \label{fig:H2Radio}
    \end{figure*}

\section{Discussion} \label{Discussion}

    \subsection{Mrk 266 SW Dominated by Rotation}
    During advanced-stage mergers, the disk of molecular gas can become decoupled from the stars due to merger-related perturbations \citep{Muller_Sanchez2018}, exhibiting chaotic dynamics with weak or absent rotational components.  While the NE nucleus appears to exhibit this behavior, the velocity map of molecular gas in the SW nucleus (\fig{fig:H2SW}) appears to be dominated by ordered rotation.  To verify this, we performed kinematic modelling of the stellar, $\text{H}_2$, and $\text{Br}\gamma$ velocity maps for the SW nucleus as well as $\text{H}_2$ and $\text{Br}\gamma$ for the NE nucleus using the software MocKinG written by Jeremy Dumoulin and Benoît Epinat \citep{MocKinG}.  The software was designed to fit kinematic models to velocity maps of high redshift galaxies by correcting for beam smearing effects, and the PyMultiNest Nested Sampling Monte Carlo implementation \citep{Buchner2014} was used to perform the fitting and estimate errors.  This software was chosen due to its ability to fit simple, physically motivated models.  This approach is ideal for subtracting only the rotating disk of the galaxy so that complex motions induced by the merger can be analyzed.  When modelling the SW stellar velocity map, the center of rotation was fixed at the AGN while all other parameters were left free.  Then, when modelling the $\text{H}_2$ and $\text{Br}\gamma$, the kinematic center coordinates, inclination, and position angle of the galaxy were fixed close to those of the stellar velocity model, allowing for variation within the errors reported in Table~\ref{tab:MocKinG-Results}.  For the NE nucleus, stellar kinematics were not extracted, so a clipped $\text{H}_2$ velocity map that removed most non-rotational motions was used to fit initial parameters.  The position angle and inclination were also measured via the morphology of the F110W HST image for comparison.
        
        The velocity curve used for all models was that for a rotating exponential disk given by equation $(\text{A}25)$ in \cite{MocKinG}:
        \begin{equation}
            $$V(r) = \frac{r}{r_0}\,
                \sqrt{\pi G \Sigma_0 r_0 \left(I_0 K_0-I_1K_1\right)}$$
        \end{equation}
        where $r_0$ is the exponential radius, $\Sigma_0$ is the central disk surface density, and $I_i$ and $K_i$ are the $i\text{th}$-order modified Bessel functions of the first and second kind.

        A simple, single-component model was chosen to allow the detection of non-rotational motions in the velocity residuals and to prevent over fitting. The exponential disk model is that of a Freeman disk in which the gravitational field is due only to the stars and is thus the simplest model for removing the rotation in the velocity field.  More complex models that account for the contribution of dark matter are not appropriate for modelling the nuclear regions for which the gravitational potential is dominated by stars.  Table~\ref{tab:MocKinG-Results} shows the results for both nuclei.  Position angles are relative to north with the counterclockwise direction positive.  Errors for each parameter were measured using the standard deviation of the posterior distributions.      

        \fig{fig:VelModel} shows the velocity models and residuals for the SW nucleus with continuum or $\text{Br}\gamma$ contours overlaid for comparison.  All residual maps indicate the SW nucleus is dominated by rotation with no large-scale chaotic motions caused by the merger.  The velocity residual map for the stars shows no structure, and appears to be a flat noise field, indicating the only kinematic component is rotation.  The $\text{H}_2$ velocity residual map shows a redshifted structure that appears to connect the peak of $\text{Br}\gamma$ emission north of the nucleus with the AGN. We interpret this structure as an inflow of molecular gas, discussed in Section \ref{ring}. The velocity residual map for $\text{Br}\gamma$ shows a conical blueshifted region east of the AGN with a velocity of \qty{-191}{\kilo\meter\per\second} that is aligned with both the $[\ion{Si}{vi}]$ spatial distribution and kinematics as well as the biconical velocity dispersion structure in $\text{Br}\gamma$, indicating that this structure is likely related to an AGN-driven outflow (see Section \ref{outflow}).  The redshifted region south of the AGN in the $\text{Br}\gamma$ velocity residual map is likely associated with stellar winds due to its spatial correlation with peaks in $\text{Br}\gamma$ flux corresponding to an increase in star formation activity.  The radial velocity is approximately \qty{50}{\kilo\meter\per\second} which is consistent with simulations of stellar winds \citep{Fichtner2023}.
        
        %Possible stellar winds associated with star formation are shown south of the AGN.  

    \subsection{Molecular Gas Inflow from Star-forming Ring in SW Nucleus} \label{ring}

        The flux map of $\text{Br}\gamma$ in \fig{fig:BrGammaSW} and the map of $\text{Br}\gamma$ equivalent width in \fig{fig:H2BrGammaRatioSW} show a clumpy, ring-like structure that is very similar to the star-forming nuclear rings studied by \citet{Boker2008}.  Therefore, we report the presence of a circumnuclear ring of star formation in the SW nucleus with a radius of approximately \qty{600}{\parsec}.  The magenta ellipse in \fig{fig:H2BrGammaRatioSW} was fit using an inclination of $\qty{72.2}{\degree}$ and a position angle of $\qty{-12.8}{\degree}$, matching the values fit by the kinematic modelling of the disk and thereby suggesting that the ring is coplanar with the disk of the galaxy.  There is no sign of the ring east of the AGN which may be caused by obscuration due to the high inclination angle (see \fig{fig:H2BrGammaRatioSW}), interference from the $[\ion{Si}{vi}]$ outflow, or clumpiness in the ring.  This is consistent with the star-forming rings studied by \citet{Boker2008}.

        $\text{H}_2$ residuals for the SW nucleus show a sub-kpc-scale redshifted region north of the AGN with a velocity of approximately \qty{70}{\kilo\meter\per\second}.  This structure flows from the region of maximum $\text{Br}\gamma$ emission to the AGN and is likely an inflow from the circumnuclear ring.  Since the residuals connect to the AGN, it is likely that the ring is the source of AGN feeding, rather than large-scale gas streamers related to the chaotic motions of the merger.

        We can make an order of magnitude estimate of the inflow rate by assuming the flow is constrained to a cylindrical pipe.  Thus,
        \begin{equation}
            \dot{M}_{\text{in}} = \rho_{\text{H}_2}\,A\,v_{\text{in}} = \frac{M_{\text{H}_2}}{l}\,v_{\text{in}}
        \end{equation}
        Where $M_{\text{H}_2}$ is the total mass of molecular gas in the inflow, $l$ is the length, and $v_{\text{in}}$ is the velocity.  Due to uncertainties in the geometry of the inflow, we do not correct for inclination and assume $v_{\text{in}}$ is the line-of-sight velocity residual after subtracting the rotating disk model, measured to be \qty[separate-uncertainty=true]{65(20)}{\kilo\meter\per\second}.  The length of the inflow is \qty[separate-uncertainty=true]{730(120)}{\parsec}.  We use the methods in \citet{Muller_Sanchez2006} and \citet{Hicks2009} to estimate $M_{\text{H}_2}$ using the luminosity of $\text{H}_2\,[1\text{--}0\,S(1)]$ within the inflow, measured to be \qty{8.4(4)e5}{\solarL}.  Assuming a ratio of ${\text{L}_{[1\text{--}0\,S(1)]}/M_{\text{H}_2}\sim \num{2.5e-4}\unit{\solarL}/\unit{\solarM}}$ with $1\sigma$ error of approximately a factor of 2 \citep{Muller_Sanchez2006}, we find a total molecular gas mass of approximately \qty{3e9}{\solarM}.  Therefore, we estimate the inflow rate to be \qty{270}{\solarM\per\year} within an order of magnitude.
        
        The accretion rate onto the AGN in Mrk 266 SW can be estimated from the bolometric luminosity of the AGN.
        \begin{equation}
            L_{\text{AGN}} = a\,\dot{M}_{\text{accretion}}\,c^2
        \end{equation}
        Where $a$ is the radiative accretion efficiency.  Theoretical studies of the radiative accretion efficiency of black holes suggest it is on the order of 0.1 \citep{a1,a2,a3}.  Additionally, \citet{DavisAndLaor} find a mean efficiency of $0.1$ via observational methods, and they report a correlation between black hole mass and efficiency such that: $$a\approx0.089\left(\frac{\text{M}_{\text{\tiny{BH}}}}{10^8\text{M}_{\odot}}\right)^{0.52}$$
        Assuming the black hole mass reported in Section~\ref{BHME} of \qty{7e7}{\solarM}, the expected accretion efficiency would be $0.074$.  However, AGN accretion efficiencies can vary widely, and \citet{Bian2003} find a decreased efficiency for Seyfert 1 galaxies with an average of $0.016$.
        
        \citet{Iwasawa2020} perform SED modeling of \textit{NuSTAR} data and find an intrinsic AGN luminosity of \qty{2.6e44}{\erg\per\second}.  Therefore, if we assume an accretion efficiency of 0.074, the accretion rate onto the AGN is approximately \qty{6.2e-2}{\solarM\per\year}, and assuming an accretion efficiency of 0.016, the accretion rate is \qty{2.9e-1}{\solarM\per\year}.  In either case, the inflow rate to the central 100 pc is $\sim\!3$ orders of magnitude larger than the accretion rate onto the SMBH.  This suggests an inefficient transport of material to the AGN and an accumulation of molecular gas in the central region of the galaxy.

        \begin{figure}
            \centering
	        % or pdf, eps, jpg if compiling using pdf latex
	        \includegraphics[width=\columnwidth]{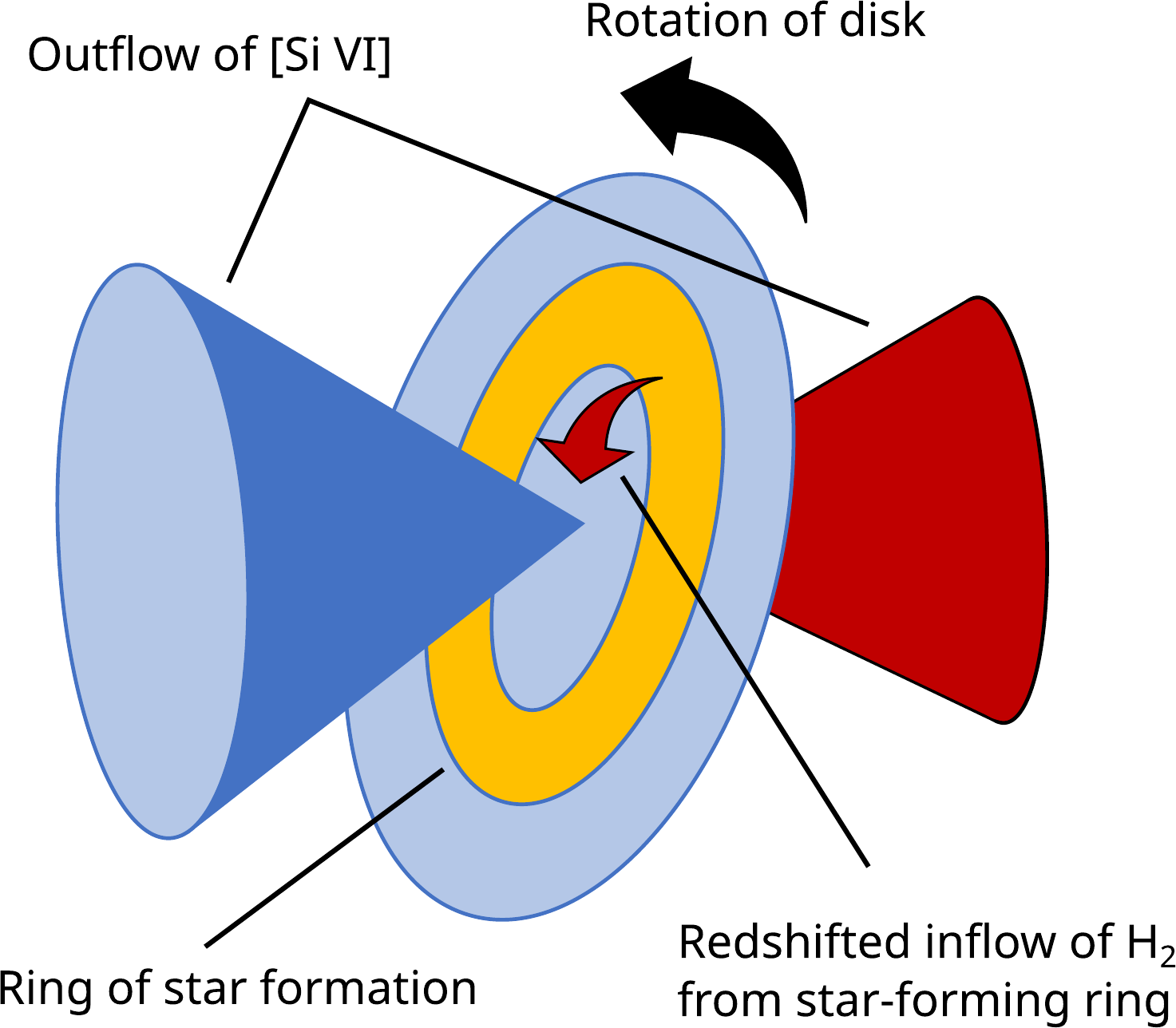}
            \caption{Graphical representation of the SW nucleus with the disk in light blue, outflow in dark blue and dark red, star-forming ring in yellow, and inflow of $\text{H}_2$ as the red arrow.}
            \label{fig:cartoonModel}
        \end{figure}

        \begin{figure*}
            \includegraphics[width=0.85\textwidth]{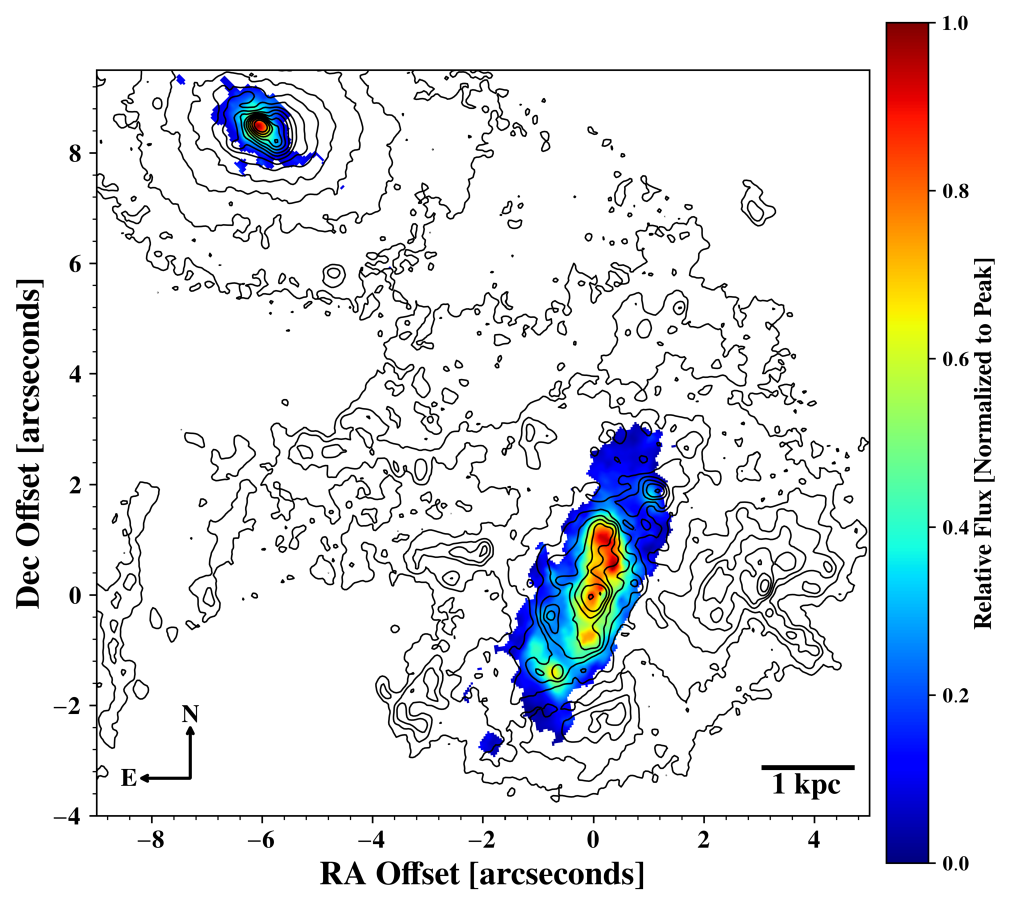}
            \caption{$\text{Br}\gamma$ flux with $\text{H}\alpha$ contours to show connection to large scale structure.  Contour lines are plotted with log scale and range from $0.05\%$ to $15\%$ of the maximum.  RA and Dec are relative to SW AGN, and the $\text{Br}\gamma$ flux for each nucleus is normalized independently.  Significant ionized winds \citep{Mazzarella2012,Ishigaki} are present at large scales despite the ordered nature of the SW nucleus.  The $\text{H}\alpha$ emission near the SW nucleus primarily traces the disk of the galaxy, and the complex structures east and west of the AGN may be the remnants of previous epochs of AGN activity \citep{Ishigaki}.}
            \label{fig:BrGammaHalpha}
        \end{figure*}

        \begin{figure*}
                        \includegraphics[width=\textwidth]{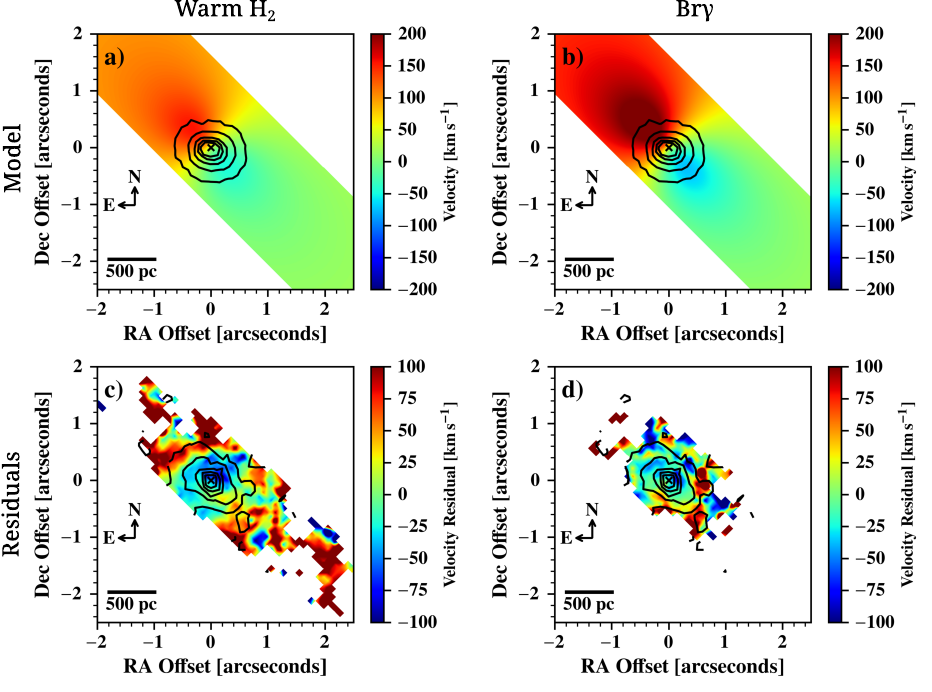}
                        \caption{Kinematic modelling results for the NE nucleus. \textit{Top:} Velocity model for $\text{H}_2$ (\textit{a}) and $\text{Br}\gamma$ (\textit{b}).  Contours are \textit{K}-band continuum ranging linearly from $1\text{\nobreakdash{--}}\num{3e-14}\,\text{erg}\,\unit{\second^{-1}\centi\meter^{-2}\micro\meter^{-1}}$. \textit{Bottom:} Velocity residuals for $\text{H}_2$ (\textit{c}) and $\text{Br}\gamma$ (\textit{d}) with contours of $\text{Br}\gamma$ relative flux ranging from $10\%$ to $80\%$ of the maximum.  There are several structures in the $\text{H}_2$ velocity residuals that are strongly decoupled from the rotation of the central region, indicating strong merger-related perturbations.}
                        \label{fig:VelModelNE}
        \end{figure*}

        \begin{figure*}
            \includegraphics[width=0.9\textwidth]{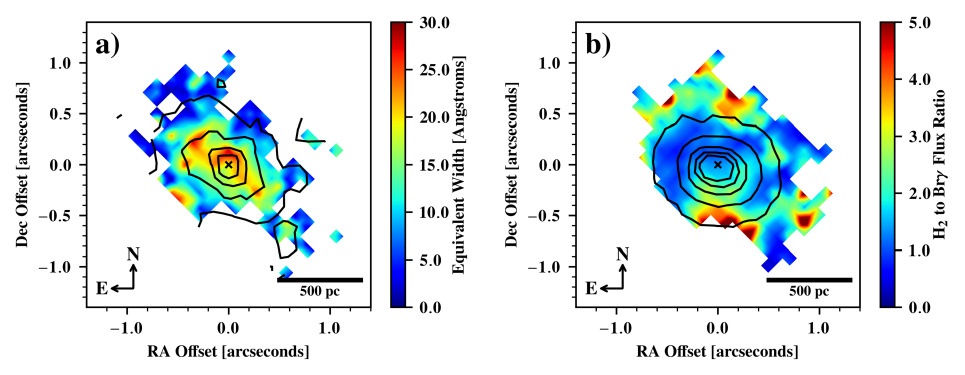}
            \caption{Equivalent width of $\text{Br}\gamma$ (\textit{a}) and $\text{H}_2$-to-$\text{Br}\gamma$ flux ratio (\textit{b}) for the NE nucleus. Contours for (\textit{a}) are $\text{Br}\gamma$ relative flux ranging linearly from $10\%$ to $80\%$ of the maximum while contours for (\textit{b}) are \textit{K}-band continuum ranging linearly from $1\text{\nobreakdash{--}}\num{3e-14}\,\text{erg}\,\unit{\second^{-1}\centi\meter^{-2}\micro\meter^{-1}}$.  The larger equivalent width near the AGN suggests that there may be less quenching of star formation in the NE nucleus, while the flux ratio map suggests the presence of shocks.}
            \label{fig:H2BrGammaRatioNE}
        \end{figure*}

    \subsection{A compact AGN-driven outflow in the SW Nucleus} \label{outflow}
        We propose that the spatial distribution and kinematics of [\ion{Si}{vi}] in Mrk 266 SW correspond to an outflow of highly-ionized gas that is both compact and primarily AGN-driven. 

        The [\ion{Si}{vi}] outflow is clearly compact ($r<1$ kpc) since the flux drops off significantly within the field of view of OSIRIS and \fig{fig:H2Radio} shows no extended radio emission perpendicular to the rotating disk that could be associated with the outflow.
                
        Four pieces of evidence suggest that the outflow is AGN-driven rather than starburst-driven. First, previous studies of the coronal line region of AGN suggest that the kinematics of this region is always dominated by radial outflow \citep{Rodriguez-Ardila2011,Muller_Sanchez2011}. Second, the flux of [Si VI] and the velocity gradient of the outflow are both perpendicular to the disk of the galaxy (\fig{fig:SiVISW}), and appear to emanate from the AGN in a geometry that resembles a sharp cone. The flux and equivalent width maps of $\text{Br}\gamma$ (\fig{fig:H2BrGammaRatioSW}), as well as the \qty{3.6}{\centi\meter} radio continuum (\fig{fig:H2Radio}), indicate that star formation is occurring within the central $\sim\! 2$ kpc of the galaxy disk. Therefore, a starburst-driven outflow would originate from a more extended region and would appear as a truncated cone or superbubble \citep{Strickland2000} with a minimum width of approximately $1-2$ \unit{\kilo\parsec}, whereas the base of the outflow in Mrk 266 SW is $\lesssim\!\qty{500}{\parsec}$.  However, the morphology alone is not sufficient to prove that the outflow is primarily AGN-driven since there could be an unresolved cluster of star formation within the central \qty{500}{\parsec} driving the outflow. The other two pieces of evidence, the dynamical timescale of the outflow and the kinematic power of the outflow, are discussed in Sections \ref{dtime} and \ref{MOR}.  \fig{fig:cartoonModel} shows a graphical representation of the SW nucleus, including the $[\ion{Si}{vi}]$ outflow, star-forming ring, and $\text{H}_2$ inflow.  Properties for the outflow are given in Table~\ref{tab:OutflowParam}.
        
        %but not near the AGN, where the outflow originates. 
        %The outflow is unlikely to be starburst-driven for the following reasons:
         %       \begin{enumerate}
          %          \item The flux map (\fig{fig:SiVISW}) is coincident with the AGN and oriented perpendicular to the disk of the host galaxy.
           %         \item The velocity gradient of the outflow is also perpendicular to the rotation of the galaxy as shown in \fig{fig:SiVISW_velProf}.
            %        \item The flux and equivalent width maps of $\text{Br}\gamma$ (\fig{fig:H2BrGammaRatioSW}), as well as the \qty{3.6}{\centi\meter} radio continuum (\fig{fig:H2Radio}), indicate that star formation is occurring within the disk but not near the AGN, where the outflow originates.
             %   \end{enumerate}

                \subsubsection{Dynamical Time of the Outflow} \label{dtime}
                
                The projected length of the outflow is \qty{0.8}{\kilo\parsec}, which corresponds to approximately \qty{0.85}{\kilo\parsec} along the outflow assuming an inclination for the outflow of \qty{20}{\degree}, consistent with a disk inclination of \qty{70}{\degree} when assuming the outflow is perpendicular to the disk.  The maximum blueshifted velocity of the outflow was measured with \fig{fig:SiVISW_velProf} and is approximately \qty{150}{\kilo\meter\per\second} or \qty{490}{\kilo\meter\per\second} along the outflow.  The dynamical time, $t_{\text{dyn}} = \text{D}/v_{\text{max}}$, is then $\sim$\qty{1.7}{\mega\year}, significantly less than the timescale of the galaxy merger (>\qty{100}{\mega\year}).  

                Since the dynamical time of the outflow is significantly shorter than the timescale of the merger, the current epoch of AGN activity in the SW nucleus must have begun during the merger. Thus, we present evidence that the SW nucleus is a young AGN triggered during an on-going merger event, despite the fact that the AGN feeding is sustained by circumnuclear processes (see Section \ref{ring}).  However, it is likely that the AGN activity in Mrk 266 SW is highly variable and episodic due to the presence of extended ionized winds to the east (see Figs.~\ref{fig:MRK266-fcc} and \ref{fig:BrGammaHalpha}) that have been shown to be seyfert ionized when analyzed with BPT diagrams \citep{Ishigaki}.  This would corroborate simulations that suggest AGN and star formation activity are episodic during the dual AGN phase \citep{Wassenhove2012,Blecha2013}. Follow-up HST observations using narrow band filters to trace [\ion{O}{iii}] emission at larger scales are required to study the outflow history and thus probe the variability of AGN activity in the SW nucleus.

                In addition, the timescale of the outflow is significantly shorter than the age of the star-forming clusters near the SW nucleus \citep[found to be 10-50 Myr by][]{Mazzarella2012}, providing further evidence that the [\ion{Si}{vi}] outflow is primarily driven by the AGN rather than star formation.

                \subsubsection{Mass Outflow Rate} \label{MOR}

                We estimate the mass outflow rate using the formulation in \cite{Muller_Sanchez2011}, given by:
                \begin{equation}
                    \dot{M}_{\text{out}} = 2\,m_p\,N_e\,v_\text{max}\,A\,f
                \end{equation}
                Where $m_p$ is the proton mass, $N_e$ is the ionized gas density assumed to be \qty{5000}{\per\centi\meter\cubed}, $v_\text{max}$ is the maximum velocity along the outflow, $A$ is the lateral surface area of the outflow cone, and $f$ is a filling factor estimated to be $0.001$.  The values for $N_e$ and $f$ were chosen based on previous studies of the coronal line region indicating ${10^3<N_e<10^4\,}\unit{\per\centi\meter\squared}$ and ${f<0.01}$ \citep{Moorwood1996,Oliva1997,Rodriguez-Ardila2006,Schneider}.  If we assume the maximum line-of-sight velocity of the outflow is \qty{150(40)}{\kilo\meter\per\second} from \fig{fig:SiVISW_velProf}, an inclination of \qty{72.2(2)}{\degree}, a projected outflow distance of \qty{0.8(1)}{\kilo\parsec}, and an opening angle of \qty{80(10)}{\degree}, ${v_\text{max}=490^{+150}_{-140}\ \unit{\kilo\meter\per\second}}$ and ${A=2.4^{+1.6}_{-1.0}\ \unit{\square\kilo\parsec}}$.  
                
                Therefore, the outflow rate is estimated to be ${300^{+340}_{-170}\ \unit{\solarM\per\year}}$.  However, the assumptions for the filling factor $f$ and $N_e$ may be off by a factor of $10$, so $\dot{M}_{\text{out}}$ should be interpreted as an order of magnitude estimate.  The kinematic power of the outflow is given by:
                \begin{equation}
                    $$\dot{E}_{\text{out}} = \frac{1}{2}\dot{M}_{\text{out}}\,v_{\text{max}}^2$$
                \end{equation}
                
                Based on the mass outflow rate, we find a kinematic power of the outflow to be ${2.3^{+6.0}_{-1.8}\!\times\!10^{36}\ \unit{\watt}}$.  This is very similar to the kinematic power of the [\ion{O}{iii}] cone in NGC 6240 that was estimated to be \qty{1.9e36}{\watt} \citep{Muller_Sanchez2018}.

                Equation 2 of \citet{GalWinds} estimates the kinematic power due to stellar activity using the Starburst99 simulations \citep{Starburst99}, assuming solar metallicity and a supernova rate of ${\sim\!0.02\left(\frac{\text{SFR}}{\unit{\solarM\per\year}}\right)\unit{\per\year}}$:
        
                $$\dot E_*=7.0\times10^{34}\left( \frac{\text{SFR}}{\unit{\solarM\per\year}} \right)\text{W}$$

                Section \ref{SFR} estimates the SFR in the central arcsecond of Mrk 266 SW to be $\qty{3.2(1)}{\solarM\per\year}$.  Using the equation above, the maximum kinematic power available from star formation within the central arcsecond is $\qty{2.2(1)e35}{W}$. Since the kinematic power of the outflow is an order of magnitude larger than that available from star formation, the maximum contribution from starbursts to the outflow is approximately 10\%. Therefore, the outflow is likely dominated by the AGN.
    
                \begin{table}
                    \centering
                    \caption{Parameters of $[\ion{Si}{vi}]$ outflow.  PA, opening angle, and projected radius were estimated with the software QFitsView.  Max redshifted velocity and max blueshifted velocity were determined by the velocity profile in Fig.~\ref{fig:SiVISW_velProf}.  Mean velocities were directly measured from the $[\ion{Si}{vi}]$ velocity map in Fig.~\ref{fig:SiVISW}.  Provided values are the observed quantities without correcting for inclination.  Details for the calculation of the timescale and mass outflow rate are in Section~\ref{Discussion}.}
                    \label{tab:OutflowParam}
                    \renewcommand{\arraystretch}{1.3}
                    \begin{tabular}{cc}
                        \hline
                        \multirow{2}{*}{Outflow Parameter} & \multirow{2}{*}{Value}   \\
                        &  \\               
                        \hline
                        PA & \qty{100\pm10}{\degree} \\
                        Opening Angle & \qty{80(10)}{\degree} \\
                        Projected Radius & \qty{0.8(1)}{\kilo\parsec} \\
                        Max Redshifted Velocity & \qty{100(50)}{\kilo\meter\per\second} \\
                        Mean Redshifted Velocity & \qty{56(3)}{\kilo\meter\per\second} \\
                        Max Blueshifted Velocity & \qty{150(40)}{\kilo\meter\per\second} \\
                        Mean Blueshifted Velocity & \qty{107(2)}{\kilo\meter\per\second} \\
                        Timescale & \qty{1.7(10)}{\mega\year} \\
                        Mass Outflow Rate & \qty{300(150)}{\solarM\per\year}\\
                        \hline
        
                    \end{tabular}
                \end{table}
    \subsection{Complex Kinematics in Mrk 266 NE}
        As opposed to the ordered rotation of the SW nucleus, the NE nucleus exhibits strong merger-related perturbations.  The velocity residuals of molecular gas in \fig{fig:VelModelNE} show a chaotic structure with many non-rotational components, indicating that the molecular gas disk in the NE nucleus has decoupled from the stellar disk.  However, we cannot perform further analysis of the decoupling due to the inability to extract stellar kinematics for the NE nucleus.  
        
        In addition, the $\text{H}_2$ to $\text{Br}\gamma$ flux ratio map in \fig{fig:H2BrGammaRatioNE} shows significantly higher ratios with more structure than the SW nucleus, suggesting the presence of shocks.  The chaotic, non-rotational motions of the molecular gas do not appear to be strongly contributing to AGN feeding since the NE nucleus is less active than the SW \citep{Mazzarella2012,Iwasawa2020}. However, AGN accretion in a chaotic environment may be highly variable and episodic, and \citet{Koss2012} finds an increase in AGN luminosity as nuclear separation decreases for kpc-scale mergers.

    \subsection{Properties of the SW and NE Nuclei}
    \subsubsection{Dynamical Mass}
        To estimate the dynamical mass of both nuclei, the following formula from \cite{Muller_Sanchez2006} was used:
        \begin{equation}
            M_{\text{dyn}} = (v_{\text{rot}}^2+3\sigma^2)r/G
        \end{equation}
        where $v_{\text{rot}}$ is the rotation velocity at the maximum radius $r$, $\sigma$ is the average velocity dispersion within $r$, and $G$ is the gravitational constant. 
        For both nuclei, the dynamical mass within a radius of \qty{500}{\parsec} of the AGN was calculated using the velocity model of the molecular gas shown in Table~\ref{tab:MocKinG-Results}.  The SW nucleus has a rotational velocity of \qty{165(5)}{\kilo\meter\per\second} at \qty{500}{\parsec} which yields approximately \qty{175}{\kilo\meter\per\second} after correcting for an inclination of \qty{70}{\degree}.  The mean velocity dispersion within \qty{500}{\parsec} is approximately \qty{90}{\kilo\meter\per\second}, resulting in a dynamical mass of \qty{6.4e9}{\solarM}.

        The NE nucleus has a rotational velocity at \qty{500}{\parsec} of \qty{157}{\kilo\meter\per\second} assuming an inclination of \qty{35}{\degree}.  The mean velocity dispersion within \qty{500}{\parsec} is \qty{155}{\kilo\meter\per\second}.  Thus, the dynamical mass is \qty{1.1e10}{\solarM}, almost double that of the SW nucleus.  While the rotational velocity is comparable to the SW nucleus, the velocity dispersion component is larger, suggesting more chaotic motions.

        \subsubsection{Black Hole Mass Estimate} \label{BHME}
        
        The $M_\text{BH}$\nobreakdash--$\sigma_*$ relation was used to estimate the black hole mass for the SW nucleus.  \citet{Medling} found that an offset must be added to the relation when applied to gas-rich mergers, and we use their formulation:
        \begin{equation}
            \label{eq:M_Sigma}
            $$\log(M_{\text{BH}}/M_{\odot}) \approx 8.32 + 5.64\,\log(\sigma_{*,\text{bulge}}/200) + 0.15 \pm 0.06$$
        \end{equation}
        We estimate the average $\sigma_*$ to be 120 km/s in the central kpc of the SW nucleus. This corresponds to a BH mass of \qty{1.7e7}{\solarM}. However, if we consider the average $\sigma_*$ in the central 500 pc of the galaxy ($\sim$\qty{170}{\kilo\meter\per\second}), this would result in $\sim$\qty{1.2e8}{\solarM}. We therefore consider an approximate black hole mass of \qty{7e7}{\solarM} for the SW nucleus with an uncertainty of a factor of 2.5. For comparison, \cite{Mazzarella2012} report a black hole mass for the SW nucleus of \qty{2.3e8}{\solarM} using the \textit{H}-band bulge luminosity scaling relation \citep{MarconiAndHunt}.  Mazzarella et al.\ use GALFIT to determine parameters of the bulge, fitting a Sersic profile ($n=4$) with an effective radius of \qty{4.92}{\kilo\parsec} and an axis ratio of $0.49$.  

        \citet{Mazzarella2012} also find a black hole mass for the NE nucleus of \qty{2.6e8}{\solarM}.  We are unable to provide a mass for the NE nucleus from the $M_\text{BH}$\nobreakdash--$\sigma_*$ relation for the NE nucleus due to the inability to extract stellar kinematics.  However, our results for the larger $M_{\text{dyn}}$ in the NE nucleus are consistent with the presence of a larger black hole.

        \subsubsection{Star Formation Rates} \label{SFR}
        \citet{Pasha2020} find a tight relation between $\text{Br}\gamma$ luminosity and SFR.  They provide two relations, one using $\text{Br}\gamma$ luminosity to estimate SFR and another using $\text{Br}\gamma$ equivalent width to estimate specific star formation rate (sSFR) which is the SFR per stellar mass:
        \begin{gather}
            \label{eq:SFR_lum}
            \log(\text{SFR}[\unit{\solarM\per\year}]) = -6.17 + 1.071\,\log(\text{L}_{\text{Br}\gamma}[\unit{\solarL}]) \\
            \log(\text{sSFR}[\unit{\per\year}]) = 1.1\,\log(\text{EW}_{\text{Br}\gamma}[\unit{\angstrom}]) - 10.63
        \end{gather}
        The intrinsic scatter for both relations is approximately ${0.1\ \text{dex}}$.

        Table~\ref{tab:EmissionLineFlux} reports the full FOV $\text{Br}\gamma$ emission line fluxes for the SW and NE nuclei as \num{24.1(25)e-15} and \qty{4.57(16)e-15}{\erg\per\second\per\centi\meter\squared}, respectively.  The luminosities are then \num{1.08(12)e7} and \qty{2.04(8)e6}{\solarL} assuming a distance of \qty{120}{\mega\parsec}.  \citet{Mazzarella2012} use mid-infrared emission line diagnostics to estimate the AGN contribution to bolometric luminosity as approximately $50\%$ with a standard deviation of $30\%$ for both nuclei.  Therefore, we estimate the SFR to be $11.0^{+9.3}_{-7.3}$ and $1.8^{+1.3}_{-1.2}\ \unit{\solarM\per\year}$ for the SW and NE nuclei, respectively.  The equivalent width of $\text{Br}\gamma$ for the full FOV is \num{40.4(85)} and \qty{16.2(9)}{\angstrom} for the SW and NE nuclei.  After accounting for the contribution from the AGN, the sSFR is ${6.4^{+10.3}_{-5.0}\!\times\!10^{-10}}$ and ${2.3^{+2.9}_{-1.7}\!\times\!10^{-10}\,\unit{\per\year}}$ for the SW and NE nuclei, respectively.  However, the relations published by \citet{Pasha2020} are for isolated galaxies without AGN in which the SFR for the entire galaxy is estimated, rather than only the nuclear region.  Therefore, these results should be interpreted as an order of magnitude estimate.
        
        \citet{Howell2010} use IRAS $\text{L}_{\text{ir}}$, GALEX FUV, and 2MASS \textit{K}-band observations to estimate a SFR of $\qty{65}{\solarM\per\year}$ and a sSFR of \qty{3.92e-10}{\per\year} for the whole Mrk 266 system.  \citet{Beaulieu2023} measure the SFR in the core of the Mrk 266 system to be \qty{15(2)}{\solarM\per\year} via $\text{H}_{\alpha}$ luminosity.  Thus, our calculations are within an order of magnitude of those previously measured with different methods.

        The ratio of the mass outflow rate of the [\ion{Si}{vi}] outflow to the SFR of the SW nucleus, $\eta$, is $\sim\!25$.  \citet{Roberts2020} study stellar outflows with the SDSS-VI/MaNGA survey and find that $\eta\!\lesssim\!5$ near the nuclear region. Therefore, it is very unlikely that the [\ion{Si}{vi}] outflow is generated primarily by star formation.  In addition, Section \ref{MOR} estimates the kinematic power that can be produced by star formation and compares it to that of the [\ion{Si}{vi}] outflow.  If it were starburst driven, only star formation within the central \qty{500}{\parsec} could contribute to the outflow.  Therefore, we must also estimate the SFR within a $1\arcsec$ diameter aperture centered on the AGN using the $\text{Br}\gamma$ flux of $\qty{3.83(10)e-15}{\erg\per\second\per\centi\meter\squared}$ reported in column 4 of Table \ref{tab:EmissionLineFlux}.  Using the same methods described above, the SFR within the central arcsecond is approximately $\qty{3.2(1)}{\solarM\per\year}$.

        \subsubsection{Merger Evolution and Large-scale Structure}
            The $\text{H}_2$ velocity map for the NE nucleus in \fig{fig:H2NE} shows a redshifted region approximately \qty{2.5}{\arcsecond} southwest of the AGN with a velocity of \qty{110}{\kilo\meter\per\second}.  This may be connected to the bridge of molecular gas between the nuclei observed by \cite{Mazzarella2012} and \cite{Beaulieu2023}.  If the SW nucleus is behind the NE, then gas would be streaming from the NE to the SW, potentially causing the increased activity in the SW nucleus.  However, to confirm this, high-resolution IFU observations with a larger field-of-view are required in order to trace the flow of gas between the two nuclei.

            \fig{fig:BrGammaHalpha} and \fig{fig:H2Radio} show the $\text{Br}\gamma$ and $\text{H}_2$ flux maps with $\text{H}\alpha$ and \qty{3.1}{\centi\meter} continuum contours \citep{Condon1991} respectively.  The $\text{H}\alpha$ image highlights that while the SW nuclear region lacks strong perturbations, there are significant ionized winds at larger scales \citep{Ishigaki}.  In the radio image, the region between the two nuclei is likely a region of shocked molecular gas at the boundary of the two colliding host galaxies \citep{Mazzarella2012}.  In addition, the SW nuclear region has radio emission aligned with the disk of the galaxy, indicating that star formation is occurring within the disk.  Conversely, the radio emission in the NE nucleus is much more compact, though extended approximately $0.4\arcsec$ to the SW.  \citet{Mazzarella2012} interprets extended \textit{B}-band emission and \qty{18}{\centi\meter} radio continuum emission in the NE nucleus as an AGN-driven outflow with a position angle of $\sim\!56\unit{\degree}$.  However, due to the complex velocity structure of the $\text{H}_2\,[1\text{--}0\,S(1)]$ emission, the inability to extract stellar kinematics for the NE nucleus with the OSIRIS data, low SNR of emission lines beyond a radius of \qty{500}{\kilo\parsec}, and the lack of high ionization lines, we are unable to disentangle perturbations directly related to the merger from those due to a possible AGN-driven outflow in the NE nucleus.  Follow-up observations with JWST NIRSpec would allow a much deeper investigation into the NE nucleus due to its higher sensitivity, lack of telluric absorption, larger spectral range, and square FOV.

\section{Summary and Conclusion} \label{five}
        We have presented high-spatial-resolution IFU data collected with Keck OSIRIS for the confirmed dual AGN Mrk 266.  Our results can be summarized as follows:
    
        \begin{enumerate}
            \item While the molecular gas disk of the NE nucleus has decoupled from the stars due to merger-related perturbations, the SW nucleus shows primarily ordered rotation with a circumnuclear ring of star formation.
            \item Velocity residuals for the SW molecular gas indicate an inflow from the ring of star formation to the AGN, suggesting that circumnuclear processes in the disk are responsible for fueling the AGN in Mrk 266 SW rather than large-scale inflows directly related to the merger.
            \item We detect a compact, primarily AGN-driven outflow of highly ionized gas in the SW nucleus. The timescale of the outflow is approximately \qty{2}{\mega\year}, significantly shorter than the timescale of the merger and star-formation.  This indicates that, while the AGN is fueled via circumnuclear processes, it is likely correlated with the large-scale processes caused by the merger.  However, follow-up observations with HST narrow band filters are required to confirm the AGN outflow history via extended [\ion{O}{iii}] emission.\
        \end{enumerate}

        \citet{Stemo} study a large sample of AGN in merging galaxies.  While they find increased activation of AGN in merging galaxies, they find no correlations between AGN luminosity and merger separation, and the AGN luminosities within merging galaxies are statistically identical to the general AGN population.  In addition, \citet{Smethurst2019} show that smooth, planar inflows from processes such as bar and spiral arm perturbations within the disk are more efficient at contributing to SMBH growth than the chaotic inflows expected from galaxy mergers where the disks are disrupted.  Our results are an ideal case study corroborating these previous results and supporting a more complex picture of the role mergers have in the SMBH-galaxy connection, in which mergers can trigger sustained AGN growth though processes commonly found in isolated AGN.  
        
        However, the merger selection process utilized by \citet{Stemo} is biased against those with small nuclear separations, and \citet{Koss2012} and \citet{Koss2018} find an increase in AGN luminosity for mergers with separations $\lesssim\!10\,\unit{kpc}$ as well as a higher probability of obscured dual AGN merger remnants.  Therefore, as the merger in Mrk 266 progresses and the molecular gas decouples from the stellar kinematics, it is possible that the dominant accretion mechanisms will change.  Indeed, the NE nucleus appears to have already begun this new phase of AGN evolution.

        These interpretations can be further supported with additional high-spatial-resolution observations of Mrk 266, including narrow filter images of the optical [\ion{O}{iii}] emission line to isolate the morphology of the outflow from perturbations due to the galaxy merger seen in $\text{H}\alpha$ at large scales.  Also, future observations with JWST NIRSpec, spanning longer wavelengths to include the other $^{12}\text{CO}$ absorption bands as well as the high-ionization line [\ion{Ca}{viii}], are required to provide more robust stellar and ionized gas kinematics with a high spatial resolution, larger FOV, and higher sensitivity.

 \section*{Acknowledgements}

 Hubble and VLA data was collected using the NASA/IPAC Extragalactic Database (NED) operated by the Jet Propulsion Laboratory, California Institute of Technology, under contract with NASA.  The images presented here were created with data from HST proposals 10592 (ACS Survey of Luminous Infrared Galaxies, PI A. Evans), 9379 (Near Ultraviolet Imaging of Seyfert Galaxies, PI H. Schmitt), and 14095 (Calibrating the Dusty Cosmos, PI G. Brammer).  

 The data presented herein were obtained at the W. M. Keck Observatory, which is operated as a scientific partnership among the California Institute of Technology, the University of California and the National Aeronautics and Space Administration. The Observatory was made possible by the generous financial support of the W. M. Keck Foundation.  The authors wish to recognize and acknowledge the very significant cultural role and reverence that the summit of Maunakea has always had within the indigenous Hawaiian community.  We are most fortunate to have the opportunity to conduct observations from this mountain.  This research has made use of the Keck Observatory Archive (KOA), which is operated by the W. M. Keck Observatory and the NASA Exoplanet Science Institute (NExScI), under contract with the National Aeronautics and Space Administration. This research has also made use of the SIMBAD database,
operated at CDS, Strasbourg, France.
 
 MR and F.M-S acknowledge support from NASA through ADAP award 80NSSC19K1096. ET acknowledges support from ANID through Millennium Science Initiative Program - NCN19\_058, CATA-BASAL - ACE210002 and FB210003, FONDECYT Regular 1200495.

 The authors would like to thank the referee for their constructive comments. MR would like to thank Joseph Mazzarella for many stimulating and enlightening conversations about Mrk 266 during the revision of this manuscript.

 \section*{Data Availability}
 The Keck/OSIRIS data are available at the Keck Observatory Archive at \url{https://koa.ipac.caltech.edu}. The HST data are available at the MAST Archive operated by the Space Telescope Science Institute at \url{https://mast.stsci.edu/portal/Mashup/Clients/Mast/Portal.html}. The VLA $3.6$ cm data from \citet{Condon1991} was obtained via the NASA/IPAC Extragalactic Database (NED) and is available at \url{https://ned.ipac.caltech.edu/uri/NED::Image/fits/1991ApJ...378...65C/NGC_5256:I:3.6cm:chy1991}. The derived data generated in this research will be shared on reasonable request to the corresponding author.

% The Acknowledgements section is not numbered. Here you can thank helpful
% colleagues, acknowledge funding agencies, telescopes and facilities used etc.

% %%%%%%%%%%%%%%%%%%%%%%%%%%%%%%%%%%%%%%%%%%%%%%%%%%
% \section*{Data Availability}

% The inclusion of a Data Availability Statement is a requirement for articles published in MNRAS.
% Data Availability Statements provide a standardised format for readers to understand the availability
% of data underlying the research results described in the article. The statement may refer to original
% data generated in the course of the study or to third-party data analysed in the article. The statement
% should describe and provide means of access, where possible, by linking to the data or providing the
% required accession numbers for the relevant databases or DOIs.

% %%%%%%%%%%%%%%%%%%%% REFERENCES %%%%%%%%%%%%%%%%%%

% % The best way to enter references is to use BibTeX:

\bibliographystyle{mnras}
\bibliography{Ruby2024a} % if your bibtex file is called example.bib

% % Alternatively you could enter them by hand, like this:
% % This method is tedious and prone to error if you have lots of references
% %\begin{thebibliography}{99}
% %\bibitem[\protect\citeauthoryear{Author}{2012}]{Author2012}
% %Author A.~N., 2013, Journal of Improbable Astronomy, 1, 1
% %\bibitem[\protect\citeauthoryear{Others}{2013}]{Others2013}
% %Others S., 2012, Journal of Interesting Stuff, 17, 198
% %\end{thebibliography}

% %%%%%%%%%%%%%%%%%%%%%%%%%%%%%%%%%%%%%%%%%%%%%%%%%%

% %%%%%%%%%%%%%%%%% APPENDICES %%%%%%%%%%%%%%%%%%%%%


%%%%%%%%%%%%%%%%%%%%%%%%%%%%%%%%%%%%%%%%%%%%%%%%%%

% Don't change these lines
\bsp	% typesetting comment
\label{lastpage}
\end{document}